\documentclass[12pt, authoryear]{elsarticle}

\usepackage{natbib}
\usepackage{hyperref}
\usepackage{latexsym}
\usepackage{graphics}
\usepackage{amsfonts}
\usepackage{enumerate}
\usepackage{graphicx}
\usepackage{epsf,boxedminipage,lscape}
\usepackage{epstopdf}
\usepackage{epsfig}
\usepackage{verbatim}
\usepackage{subfig}
\usepackage{caption}
\usepackage{footnote}
\usepackage{lscape}
\usepackage{hhline}
\usepackage{array}
\usepackage{graphicx}
\usepackage{setspace}
\usepackage{amssymb}
\usepackage{amsmath}
\usepackage{amsfonts}
\usepackage{times}
\usepackage{color}
\usepackage{amsthm}
\usepackage{indentfirst}
\usepackage{epsfig}
\usepackage{lscape,ctable,rotating}
\usepackage{enumitem}
\setlist{nolistsep}
\usepackage[left=1in,top=1in,right=1in,foot=1in]{geometry}

\newtheorem{definition}{Definition}[section]

\newtheorem{exam}[definition]{Example}

\newtheorem{rem}{Remark}[section]

\renewcommand{\P}{\mathbb{P}}
\newcommand{\Q}{\mathbb{Q}}
\newcommand{\var}{\textrm{\rm var}}

\newcommand{\levy}{{L\'evy}}

\makeatletter
\newcommand{\vast}{\bBigg@{4}}
\newcommand{\Vast}{\bBigg@{5}}
\makeatother

\hypersetup{pdfauthor=author}
\begin{document}

\begin{frontmatter}

\title{Deep Calibration With Artificial Neural Network: A Performance Comparison on Option Pricing Models}

\author[1]{Young Shin Kim}
\ead{aaron.kim@stonybrook.edu}
\author[2]{Hyangju Kim\corref{cor1}}
\ead{hyangju.kim@gmail.com}
\author[3]{Jaehyung Choi\corref{cor2}}
\ead{jj.jaehyung.choi@gmail.com}

\address[1]{College of Business, Stony Brook University, New York, USA}
\address[2]{Citibank, Inc. New York, USA}
\address[3]{Goldman Sachs \& Co., New York, USA}

\cortext[cor1]{Disclaimer: The views, information, or opinions expressed in this publication are solely those of Hyangju Kim and do not represent those of Citigroup, its respective affiliates, or employees.}
\cortext[cor2]{Disclaimer: The opinions and statements expressed in this article are those of the authors. None of Goldman Sachs or their respective affiliates offer, or are otherwise the source of, any of the opinions or
statements set out in this article (which may be different to views or opinions held or expressed by or within Goldman Sachs). The content of this article is for information purposes only and is not investment advice (or advice of any other kind). None of the authors, Goldman Sachs, or their respective affiliates, officers, employees, or representatives accepts any liability whatsoever in connection with any of the content of this article or for any action or inaction of any person taken in reliance upon such content (or any part thereof).}

\begin{abstract}
    This paper explores Artificial Neural Network (ANN) as a model-free solution for a calibration algorithm of option pricing models. We construct ANNs to calibrate parameters for two well-known GARCH-type option pricing models: Duan's GARCH and the classical tempered stable GARCH that significantly improve upon the limitation of the Black-Scholes model but have suffered from computation complexity. To mitigate this technical difficulty, we train ANNs with a dataset generated by Monte Carlo Simulation (MCS) method and apply them to calibrate optimal parameters. The performance results indicate that the ANN approach consistently outperforms MCS and takes advantage of faster computation times once trained. The Greeks of options are also discussed.
    
    
\end{abstract}

\end{frontmatter}        

\section{Introduction}
\label{Introduction}
    Option pricing has been a centric topic in quantitative finance for the past few decades, with remarkable growth in option market. Since the seminal work of \cite{BlackScholes:1973} and \cite{merton1973theory}, the Black-Scholes model has remained the most fundamental model for option pricing. However, its restrictive assumptions, such as constant volatility or Geometric Brownian Motion (GBM), have been criticized for not reflecting the empirical characteristics of financial markets.

    Many subsequent models have since been proposed to relax the assumptions of the Black-Scholes model. One successful approach is employing stochastic volatility under the generalized autoregressive conditional heteroskedastic (GARCH) framework. The early attempt was introduced by \cite{engle1992implied} focusing on implied conditional volatilities. Subsequently, \cite{Duan:1995} developed a more rigorous framework of the GARCH option pricing model using the locally risk-neutral valuation relationship that one-period ahead conditional variance remains constant under both the risk-neutral measure and the physical measure. The model extended later to the jump-diffusion model in \cite{Duan_et_al:2004} and \cite{Duan_et_al:2006}. Duan's GARCH model assumes that the residuals follow a normal distribution, however, empirical evidence shows that the assumption of a normally distributed residual in the model does not properly describe asset return dynamics (\cite{duan1999conditionally} and \cite{menn2009smoothly}).

    Another approach, known as the \levy~distribution, has been developed allowing for jumps, skewness, kurtosis, and heavy tails in underlying distribution to overcome the flaws of the GBM assumption. It is also referred to as a stable distribution, and its one well-known subclass is the classical tempered stable (CTS) distribution.\footnote{The other subclasses are the classical tempered stable process, generalized classical tempered stable process, modified tempered stable process, normal tempered stable process, Kim-Rachev tempered stable process, and rapidly decreasing tempered stable process.} The CTS distribution has been researched in several directions such as portfolio management (\cite{tsuchida2012mean,beck2013empirical,georgiev2015periodic,anand2016foster,choi2021diversified}) and momentum strategy (\cite{choi2015reward}). 
    
    As a consolidation of these two approaches in the context of option pricing, \cite{Kim_et_al:2008b}, \cite{Kim_et_al:2010:JBF} and \cite{KimRhoDouady:2022} enhanced Duan's GARCH model by incorporating the classical tempered stable (CTS) distribution \footnote{The CTS distribution has been studied under different names including the \textit{truncated \levy~flight} by \cite{Koponen:1995}, the \textit{tempered stable} by \cite{BarndorffNielsenLevendorskii:2001}, \cite{BarndorffNielsenShephard:2001} and \cite{Cont_Tankov:2004}, the \textit{KoBoL} distribution by \cite{Boyarchenko_Levendorskii:2000}, and the \textit{CGMY} by \cite{CGMY:2002}. The KR distribution of \cite{Kim_et_al:2008a} is an extension of the CTS distribution. \cite{Rosinski:2007} generalized CTS distribution, referred to as the tempered stable distribution.} and referred to it as the CTS-GARCH model. The CTS-GARCH model, which takes into account the non-normality in its innovation process, is considered one of the most advanced option pricing models as it addresses two key limitations of the Black-Scholes model simultaneously (\cite{kim2010tempered}). 

    However, these models are mostly high-dimensional, and this becomes problematic in the stage of parameter calibration; parameters need to be tuned, so as to find the best model output closest to the market price. Hence the Black-Scholes model still has been a practically useful model even with the superior performance of multidimensional models. 

    Recently, machine learning has shed a different light on the curse of dimensionality as a nonparametric or model-free solution. This is a rapidly evolving area with the maturation of computation power and technical advances in algorithms.

    One of the main pillars of machine learning is Artificial Neural Network (ANN). ANN, also referred to as neural network or multilayer perceptions, has been introduced by \cite{mcculloch1943logical}. ANNs are composed of the input layer, multiple hidden layers, and the output layer where each hidden layer performs a vector-to-vector or vector-to-scalar calculation to make the best output layer approximation. The parallel structure of hidden layers allows using multiprocessors, hence ANN is considerably faster than traditional algorithms at computation speed. With this advantage, ANN has been developed rapidly and different types of ANN have been proposed such as multilayer perceptron neural network, convolutional neural network, radial basis function neural network, recurrent neural network, modular neural network, and so on. 

    Many studies in finance also sought to find its ANN applications. In particular, ANN receives attention as a promising alternative for chronic computational difficulties in option pricing models as it is built upon multidimensional nonlinear models. The early attempts were applying ANN as a functional approximator for the Black-Scholes formula suggested by \cite{malliaris1993neural} and \cite{hutchinson1994nonparametric}. Since then, over a hundred papers have studied ANN for option pricing with various parameter inputs and performance measures (\cite{ruf2019neural}). 
    
    While most of the studies utilize ANN to estimate the option price based on parameters from the market data and measure their performance by out-of-sample tests, a different approach was proposed by \cite{bayer2019deep} and \cite{alaya2021deep}, recently. Rather than focusing on the approximation of the pricing formula, ANN is applied to the calibration stage and finds model parameters in this approach. Since training sets are generated by Monte Carlo Simulation (MCS) method, hence a more sound dataset avoiding market incompleteness is available. Utilizing enough large datasets is another advantage.  

    This paper also focuses on ANN as an alternative calibration method. We generate training sets having 100,000 simulated vectors of parameters and prices using the MCS method for each of Duan's GARCH and CTS-GARCH models, and then train ANNs that consist of three hidden layers with twenty nodes for each layer. Specifically, a feedforward neural network is used, which does not include any cycles or loops. The trained ANNs are utilized in the calibration for S\&P 500 index call and put option prices from every second Wednesday of each month between June 2021 and May 2022. We find that this approach not only presents a superior performance to the previous MCS method but performs remarkably faster. Additionally, the option Greeks are also computed by using the ANN method.

    The remainder of this paper is organized as follows. We discuss the GARCH option pricing models in Section \ref{Preliminaries}. Section \ref{Artificial Neural Network} presents the construction and training methods for ANN. The performance of ANN in terms of calibration is investigated using empirical data in Section \ref{Calibration}. Finally, Section \ref{Conclusion} provides our conclusions.

\section{Preliminaries}
\label{Preliminaries}
    In this section, we revisit Duan's GARCH and CTS-GARCH option pricing models that feature \textit{infinitely divisible} innovations. An infinitely divisible random variable is a random variable that can be represented as an infinite sum of independent, identically distributed random variables. This concept plays a significant role in the study of stable distributions and option pricing models.
    
    We begin by reviewing the CTS distribution as an example of an infinitely divisible distribution, followed by an overview of Duan's GARCH and CTS-GARCH option pricing models. Lastly, we outline European call and put option pricing using the MCS method for both GARCH option pricing models.
   
\subsection{CTS Distribution}
    $X$ is referred to as the \textit{Classical Tempered Stable distributed random variable} and denoted by $X\sim \textup{CTS}(\alpha$, $C$, $\lambda_+$, $\lambda_-$, $m)$ (\cite{rachev2011financial} and \cite{kim2010tempered}) if the characteristic function of the distribution is expressed as
\label{CTS Distribution}
    \begin{align}
    \begin{split}
        &E[e^{iuX}] = \phi_\textup{CTS}(u; \alpha, C, \lambda_+, \lambda_-, m) \\ 
        & = \exp\left((m-C\Gamma(1-\alpha)(\lambda_+^{\alpha-1}-\lambda_-^{\alpha-1})) iu -C\Gamma(-\alpha)\left((\lambda_+-iu)^{\alpha}-\lambda_+^{\alpha}+(\lambda_-+iu)^{\alpha}-\lambda_-^{\alpha}\right)\right)
    \end{split}
    \end{align} 
    where $C_+, C_-, \lambda_+, \lambda_-$ are positive, $0<\alpha<2$, $m\in\mathbb{R}$ and $\Gamma$ is the gamma function.
    
    If we substitute $C=(\Gamma(2-\alpha)(\lambda_+^{\alpha-2}+\lambda_-^{\alpha-2}))^{-1}$ and $m=0$ for $Z\sim \textup{CTS}(\alpha$, $C$, $\lambda_+$, $\lambda_-$, $m)$, then we have $E[Z]=0$ and $\var(Z)=1$. In this case, the random variable $Z$ is called the \textit{standard CTS distribution}, and $Z\sim \textup{stdCTS}(\alpha$,  $\lambda_+$, $\lambda_-)$. The characteristic function of $Z$ is given by
    \begin{align}
    \begin{split}
        &E[e^{iuZ}] = \phi_\textup{stdCTS}(u; \alpha, \lambda_+, \lambda_-) \\
        &=\exp\left(\frac{\lambda_+^{\alpha-1}-\lambda_-^{\alpha-1}}{(\alpha-1)(\lambda_+^{\alpha-2}+\lambda_-^{\alpha-2})}iu
        +\frac{(\lambda_+-iu)^{\alpha}-\lambda_+^{\alpha}+(\lambda_-+iu)^{\alpha}-\lambda_-^{\alpha}}{\alpha(\alpha-1)(\lambda_+^{\alpha-2}+\lambda_-^{\alpha-2})}\right).
    \end{split}
    \end{align}
    We denote the function of log-Laplace transform of $Z$ as follows (\cite{kim2010tempered}):
    \begin{align}
    \label{eq:LaplaceTransformStdCTS}
        l(x) &:= \log\left(\phi_\textup{stdCTS}(-ix; \alpha, \lambda_+, \lambda_-)\right) \\
        \nonumber
        &=\frac{x(\lambda_+^{\alpha-1}-\lambda_-^{\alpha-1})}{(\alpha-1)(\lambda_+^{\alpha-2}+\lambda_-^{\alpha-2})}
        +\frac{(\lambda_+-x)^{\alpha}-\lambda_+^{\alpha}+(\lambda_-+x)^{\alpha}-\lambda_-^{\alpha}}{\alpha(\alpha-1)(\lambda_+^{\alpha-2}+\lambda_-^{\alpha-2})}.
    \end{align}

\subsection{Duan's GARCH and CTS-GARCH Option Pricing Models}
\label{Duan's GARCH and CTS-GARCH Option Pricing Models}
    Let $(S_t)_{t\in\{0,1,\cdots, T^*\}}$ be the underlying asset price process and $(y_t)_{t\in\{0,1,2,\cdots, T^*\}}$ be the underlying asset log return process, where $y_t = \log(\frac{S_t}{S_{t-1}})$ with $y_0=0$, and $T^*<\infty$ in the time horizon. Under the physical measure $\P=\bigoplus_{t=1}^{T^*}\mathcal P_t$, $(y_t)_{t\in\{0,1,2,\cdots, T^*\}}$ is supposed to follow the GARCH model:
    \begin{equation}
    \label{eq:GARCHmodel}
        \begin{cases}
            y_{t+1}= \mu_{t+1} + \sigma_{t+1}\epsilon_{t+1}\\
            \sigma_{t+1}^2 = \kappa + \xi \sigma_{t}^2\epsilon_{t}^2 + \zeta\sigma_{t}^2
        \end{cases}
        \text{ for } t\in\{0, 1,2,\cdots, T^*-1\},
    \end{equation}
    where $\mu_{t+1}$ is the daily expected return and $\epsilon_{t+1}$ follows an infinitely divisible distribution. $S_0$, $\epsilon_{0}$ and $\sigma_0$ are real constants with $\epsilon_{0}=0$. The GARCH model involves parameters with $\kappa$, $\xi$, $\zeta$ where $\xi+\zeta<1$. 
    
    We next define $(R_t)_{t\in\{1,2,\cdots, T^*\}}$ and $(d_t)_{t\in\{1,2,\cdots, T^*\}}$ as the sequences of the daily risk-free rate of return and daily dividend rate of the underlying, respectively. Then, there is a risk-neutral measure $\Q=\bigoplus_{t=1}^{T^*}\mathcal Q_t$ such that:
    \begin{itemize}
        \item 
            $\eta_{t+1} = \theta_{t+1} + \epsilon_{t+t},$\\
            where $\theta_{t+1} = \frac{\mu_{t+1}-R_{t+1}+d_{t+1}+w_{t+1}}{\sigma_{t+1}}$ is the market price of risk and $\omega_{t+1}=\log E_\Q\left[e^{\sigma_{t+1}\epsilon_{t+1}}\right]$.
        \item $\eta_{t+1}$ is also infinitely divisible under the measure $\Q$.
    \end{itemize}
    By applying change of measures to Eq. \eqref{eq:GARCHmodel}, we obtain the risk-neutral price process under $\Q$ as
    \begin{align}
        \begin{cases}
            y_{t+1} = R_{t+1}-d_{t+1}-\omega_{t+1}+\sigma_{t+1}\eta_{t+1}\\
            \sigma_{t+1}^2 = \kappa+\xi \sigma_{t}^2(\eta_{t}-\theta_t)^2 + \zeta\sigma_{t}^2
        \end{cases}
        \text{ for } t\in\{0, 1,2,\cdots, T^*-1\}
    \end{align}
    with $\xi+\zeta<1$. To simplify the condition $\xi+\zeta<1$, we define two parameters $\psi$ and $\gamma$ as $\psi=\frac{\xi}{\zeta}$ and $\gamma=\xi+\zeta$, respectively. Then we have
    \begin{align}
        \sigma_t^2 = \kappa + \frac{\gamma}{\psi+1} \left(\psi\sigma_{t-1}^2(\eta_{t-1}-\theta_t)^2 + \sigma_{t-1}^2\right),
    \end{align}
    where $\kappa, \psi, \gamma>0$, $\eta_0=0$ and $\sigma_0>0$.
    
    Under the risk-neutral measure $\Q$, the underlying asset price is defined as $S_t = S_0e^{\sum_{j=1}^t y_j}$ for $t\in\{1$, $2$, $\cdots$, $T-1$, $T$, $\cdots$, $T^*\}$. The European option with a payoff function $H(S(T))$ at the maturity $T$ with $t\le T\le T^*$ is given by
    \begin{align}
        E_\Q\left[e^{-\sum_{j=t+1}^T R_j}H(S(T)) \biggm| \mathcal F_t\right] = E_\Q\left[e^{-\sum_{j=t+1}^T R_j}H(S_te^{\sum_{j=t+1}^T y_j})\biggm| \mathcal F_t\right].
    \end{align}
    For example, European vanilla call and put price with strike price $K$ and time to maturity $T$ at time $t=0$ are 
    \begin{align}
    \label{eq:CallPrice}
        \textup{(Call)} = E_\Q\left[e^{-\sum_{j=1}^T R_j}\max\{S_0e^{\sum_{j=1}^T y_j}-K,0\}\right]
    \end{align}
    and
    \begin{align}
    \label{eq:PutPrice}
        \textup{(Put)}=E_\Q\left[e^{-\sum_{j=1}^T R_j}\max\{K-S_0e^{\sum_{j=1}^T y_j},0\}\right],
    \end{align}
    respectively. 
    
    Moreover, let $m$ be the moneyness defined as $m=\frac{K}{S_0e^{\sum_{t=1}^T R_t}}$, then we have
    \begin{align}
        \textup{(Call)} = S_0 V_C \text{ and } \textup{(Put)} = S_0 V_P,
    \end{align}
    where
    \begin{align}
        V_C = E_\Q\left[\max\left\{\exp\left(\sum_{t=1}^T -d_t-w_t+\sigma_t \eta_t\right)-m,0\right\}\right]
    \end{align}
    and
    \begin{align}
        V_P = E_\Q\left[\max\left\{m-\exp\left(\sum_{t=1}^T -d_t-w_t+\sigma_t \eta_t\right),0\right\}\right].
    \end{align}
    
    Now, we present two popular examples of the GARCH option pricing model with infinitely divisible innovations:
    \begin{itemize}
        \item 
            \textit{Duan's GARCH Model} (\cite{Duan:1995}): If we assume that $\eta_{t}$'s follow the standard Gaussian distribution, which is infinitely divisible, we obtain the Duan's GARCH option pricing model where $w_t = \frac{\sigma_t^2}{2}$.
        \item 
            \textit{CTS-GARCH Model} (\cite{Kim_et_al:2010:JBF}):
            By assuming that $\eta_{t}$'s follow the standard CTS distribution, which is also infinitely divisible, we refer to the GARCH model as the CTS-GARCH option pricing model. Specifically, when we set $\eta_{t}\sim \textup{stdCTS}(\alpha, \lambda_+, \lambda_-)$ for all $t\in\{1,2,\cdots \}$ under the measure $\Q$, we have 
            \begin{align}
                w_t = \log\left(\phi_{\textup{stdCTS}}(-i\sigma_{t}; \alpha, \lambda_+, \lambda_-)\right)=l(\sigma_t)
            \end{align} 
            by Eq. \eqref{eq:LaplaceTransformStdCTS}. 
    \end{itemize}

\subsection{European Call and Put Option Pricing with the MCS Method}
    To simplify the model, we make the following assumptions for the remainder of this paper: the daily risk-free return $R_t$ is a constant $R$, $\theta_t$ is a constant $\theta$, and $d_t = 0$. We define $t\in\{1$, $2$, $\cdots$, $T$, $\cdots$, $T^*\}$ for $0<T\le T^*$ as a set of the time steps, where one step represents one business day. We also assume that there are 250 business days in a year, and we define the annual risk-free return as $r = 250\cdot R$. The year-fraction time is denoted by $\tau = \frac{T}{250}$, and we define $m$ as the moneyness, where $m=\frac{Ke^{-RT}}{S_0}=\frac{Ke^{-r\tau}}{S_0}$.
    
    Based on these assumptions, we can generate a set of infinitely divisible random numbers $\{\eta_{t,n}:$ $t\in\{1$,$2$,$\cdots$, $T^*\}$, $n\in\{1$, $2$, $\cdots$, $N\}\}$ using the MCS method. Next, we apply the GARCH model to obtain the set of volatility $\{\sigma_{t,n}:$ $t\in\{1$, $2$, $\cdots$, $T^*\}$, $n\in\{1$, $2$, $\cdots$, $N\}\}$ defined as 
    \begin{align}
        \sigma_{t,n} = \sqrt{\kappa + \frac{\gamma}{\psi+1} \left(\psi\sigma_{t-1, n}^2(\eta_{t-1}-\theta)^2 + \sigma_{t-1,n}^2\right)},
    \end{align}
    where $\kappa, \psi, \gamma>0$, $\eta_0=0$ and $\sigma_{0,n}=\sigma_0$ for a constant $\sigma_0$. 
    
    For a given time to maturity $T\le T^*$ and a moneyness $m$, we approximate $V_C$ and $V_P$ as follows:
    \begin{align}
        \label{eq:ID_MCS_Call}
        V_C \approx \hat V_C = \frac{1}{N}\sum_{n=1}^N
        \max\left\{\exp\left(\sum_{t=1}^{250\tau} -w_{t,n}+\sigma_{t,n} \eta_{t,n}\right)-m,0\right\}
    \end{align}
    and
    \begin{align}
    \label{eq:ID_MCS_Put}
        V_P \approx \hat V_P =  \frac{1}{N}\sum_{n=1}^N
        \max\left\{m-\exp\left(\sum_{t=1}^{250\tau} -w_{t,n}+\sigma_{t,n} \eta_{t,n}\right),0\right\}.
    \end{align}
    
    From \eqref{eq:ID_MCS_Call} and \eqref{eq:ID_MCS_Put}, call and put option values in Duan's GARCH and CTS-GARCH models can be calculated as follows:
    \begin{itemize}
        \item
            \textit{Duan's GARCH model:} We generate a set of standard Gaussian random numbers $\{\eta_{t,n}:$ $t\in\{1$,$2$,$\cdots$,$T^*\}$, $n\in\{1$,$2$,$\cdots$,$N\}\}$ and set  $w_{t,n} = \frac{\sigma_{t,n}^2}{2}$. Then we obtain call and put option prices of Duan's GARCH model. In this case, we denote \eqref{eq:ID_MCS_Call} and \eqref{eq:ID_MCS_Put} as
            \begin{align}
                V_C^{Duan}(m, \tau; \kappa, \psi, \gamma, \theta, \sigma_0)=\hat V_C
                ~~~\text{ and }~~~
                V_P^{Duan}(m, \tau; \kappa, \psi, \gamma, \theta, \sigma_0)=\hat V_P,
            \end{align}
            respectively. The call and put option prices can be approximated using MCS as
            \begin{align}
                \textup{Call}^{Duan}(S_0, K, \tau, r) \approx S_0V_C^{Duan}(m, \tau; \kappa, \psi, \gamma, \theta, \sigma_0)
            \end{align}
            and
            \begin{align}
                \textup{Put}^{Duan}(S_0, K, \tau, r) \approx S_0 V_P^{Duan}(m, \tau; \kappa, \psi, \gamma, \theta, \sigma_0).
            \end{align}
        \item
            \textit{CTS-GARCH model:}
            We simulate a set of standard CTS random numbers $\{\eta_{t,n}:$ $t \in\{1$,$2$,$\cdots$,$T^*\}$, $n\in\{1$,$2$,$\cdots$,$N\}\}$ with parameters $(\alpha$, $\lambda_+$, $\lambda_-)$. By setting  $w_{t,n} = l(\sigma_{t,n})$, we obtain call and put option prices of the CTS-GARCH model. Specifically, we use \eqref{eq:ID_MCS_Call} and \eqref{eq:ID_MCS_Put} to denote the call and put option values as 
            \begin{align}
                V_C^{CTS}(m, \tau; \kappa, \psi, \gamma, \theta, \sigma_0, \alpha, \lambda_+, \lambda_-)=\hat V_C
                 ~~~\text{ and }~~~V_P^{CTS}(m, \tau; \kappa, \psi, \gamma, \theta, \sigma_0, \alpha, \lambda_+, \lambda_-)=\hat V_P,
            \end{align}
            respectively. Using MCS, we can obtain call and put option prices under the CTS-GARCH model such that
            \begin{align}
                \textup{Call}^{CTS}(S_0, K, \tau, r) \approx S_0 V_C^{CTS}(m, \tau; \kappa, \psi, \gamma, \theta, \sigma_0,\alpha, \lambda_+, \lambda_-)
            \end{align}
            and
            \begin{align}
                \textup{Put}^{CTS}(S_0, K, \tau, r) \approx S_0 V_P^{CTS}(m, \tau; \kappa, \psi, \gamma, \theta, \sigma_0, \alpha, \lambda_+, \lambda_-).
            \end{align}   
    \end{itemize}

\section{Artificial Neural Network}
\label{Artificial Neural Network}
    In this section, we construct two ANNs to calculate call and put option prices under Duan's GARCH model and CTS-GARCH model.
    More precisely, we design multi-layer ANNs to generate similar results as the function values of $V_C^{Duan}$, $V_P^{Duan}$, $V_C^{CTS}$, and $V_P^{CTS}$. To facilitate the training, we take the logarithm of these four functions and define new functions as follows:
    \begin{align}
        v_C^{Duan}(m, \tau; \kappa, \psi, \gamma, \theta, \sigma_0, \alpha, \lambda_+, \lambda_-) &= \log \left(
        V_C^{Duan}(m, \tau; \kappa, \psi, \gamma, \theta, \sigma_0, \alpha, \lambda_+, \lambda_-)\right)\\
        v_P^{Duan}(m, \tau; \kappa, \psi, \gamma, \theta, \sigma_0, \alpha, \lambda_+, \lambda_-) &= \log \left(
        V_P^{Duan}(m, \tau; \kappa, \psi, \gamma, \theta, \sigma_0, \alpha, \lambda_+, \lambda_-)\right)\\
        v_C^{CTS}(m, \tau; \kappa, \psi, \gamma, \theta, \sigma_0, \alpha, \lambda_+, \lambda_-) &= \log \left(
        V_C^{CTS}(m, \tau;\kappa, \psi, \gamma, \theta, \sigma_0, \alpha, \lambda_+, \lambda_-)\right)\\
        v_P^{CTS}(m, \tau; \kappa, \psi, \gamma, \theta, \sigma_0, \alpha, \lambda_+, \lambda_-) &= \log \left(
        V_P^{CTS}(m, \tau; \kappa, \psi, \gamma, \theta, \sigma_0, \alpha, \lambda_+, \lambda_-)\right)
    \end{align}
    
\subsection{Generating Training Set}
\label{Generating Training Set}
    We first generate training sets for the four functions $v_C^{Duan}$, $v_P^{Duan}$, $v_C^{CTS}$, and $v_P^{CTS}$ using the MCS method as explained in the previous section. Since $v_C^{Duan}$ and $v_P^{Duan}$ have seven input parameters ($m$, $\tau$, $\kappa$, $\psi$, $\gamma$, $\theta$, $\sigma_0$), we consider seven nodes in the input layer. For $v_C^{CTS}$, and $v_P^{CTS}$, three additional input parameters ($\alpha$, $\lambda_+$ $\lambda_-$) are needed, in addition to the seven input parameters for $v_C^{Duan}$ and $v_P^{Duan}$. The range of the input parameters can be found in Table \ref{table:InputParameterRanges}. More details on training set generation process for each model are as follows: 

    \begin{itemize}
        \item
            \textit{Duan's GARCH Model:} We generate 100,000 seven-dimensional uniformly distributed random vectors for ($m$, $\tau$, $\kappa$, $\psi$, $\gamma$, $\lambda$, $\sigma_0$) with the boundary specified in Table \ref{table:InputParameterRanges}. To avoid the clustering of the random vector, we use the Halton algorithm (\cite{Halton:1964}). Then we calculate 100,000 of $v_C^{Duan}$ and $v_P^{Duan}$, respectively, using MCS. In MCS, we use 20,000 sample paths based on the Duan's GARCH model with the seven-dimensional model parameters.  

        \item
            \textit{CTS-GARCH Model:} We generate 100,000 ten-dimensional uniformly distributed random vectors for ($m$, $\tau$, $\kappa$, $\psi$, $\gamma$, $\theta$, $\sigma_0$, $\alpha$, $\lambda_+$, $\lambda_-$) with the boundary specified in Table \ref{table:InputParameterRanges}. Random numbers of $\lambda_+$ and $\lambda_-$ are generated by $\lambda_+=\tan(\frac{u_1\pi}{2})+0.1$ and $\lambda_-=\lambda_+=\tan(\frac{u_2\pi}{2})+0.1$, respectively, for uniform random numbers $u_1,u_2\in (0,1)$. We also use the Halton algorithm to generate uniform random vectors, as we did in the case of Duan's GARCH model. Then we calculate 100,000 of $v_C^{CTS}$ and $v_P^{CTS}$, respectively, using the MCS. In the MCS, we use 20,000 sample paths based on the CTS-GARCH model with the ten-dimensional model parameters.  
    \end{itemize}

    \begin{table}\centering
    \begin{tabular}{ccc}
    Parameter & Lower Bound & Upper Bound \\
    \hline
     $m$& $0.5$ & $1.5$ \\
     $\tau$& $0.4$ & $1$\\
     $\kappa$& $0$ & $1\cdot 10^{-5}$\\
     $\psi$& $0.1$ & $0.4$ \\
     $\gamma$& $0.5$ & $0.9999$\\
     $\theta$& $0$ & $0.8$\\
     $\sigma_0$& $1\cdot 10^{-6}$ & $0.04$\\
     \hline
     $\alpha$& $0.01$ & $1.999$\\
     $\lambda_+$&  $0.1$ & $\infty$\\
     $\lambda_-$&  $0.1$ & $\infty$\\
     \hline
    \end{tabular}
    \caption{\label{table:InputParameterRanges}The range of input parameters}
    \end{table}

\subsection{Training Multi-Layer ANNs}
\label{Training Multi-Layer ANNs}
\begin{figure}
        \includegraphics[width=\textwidth]{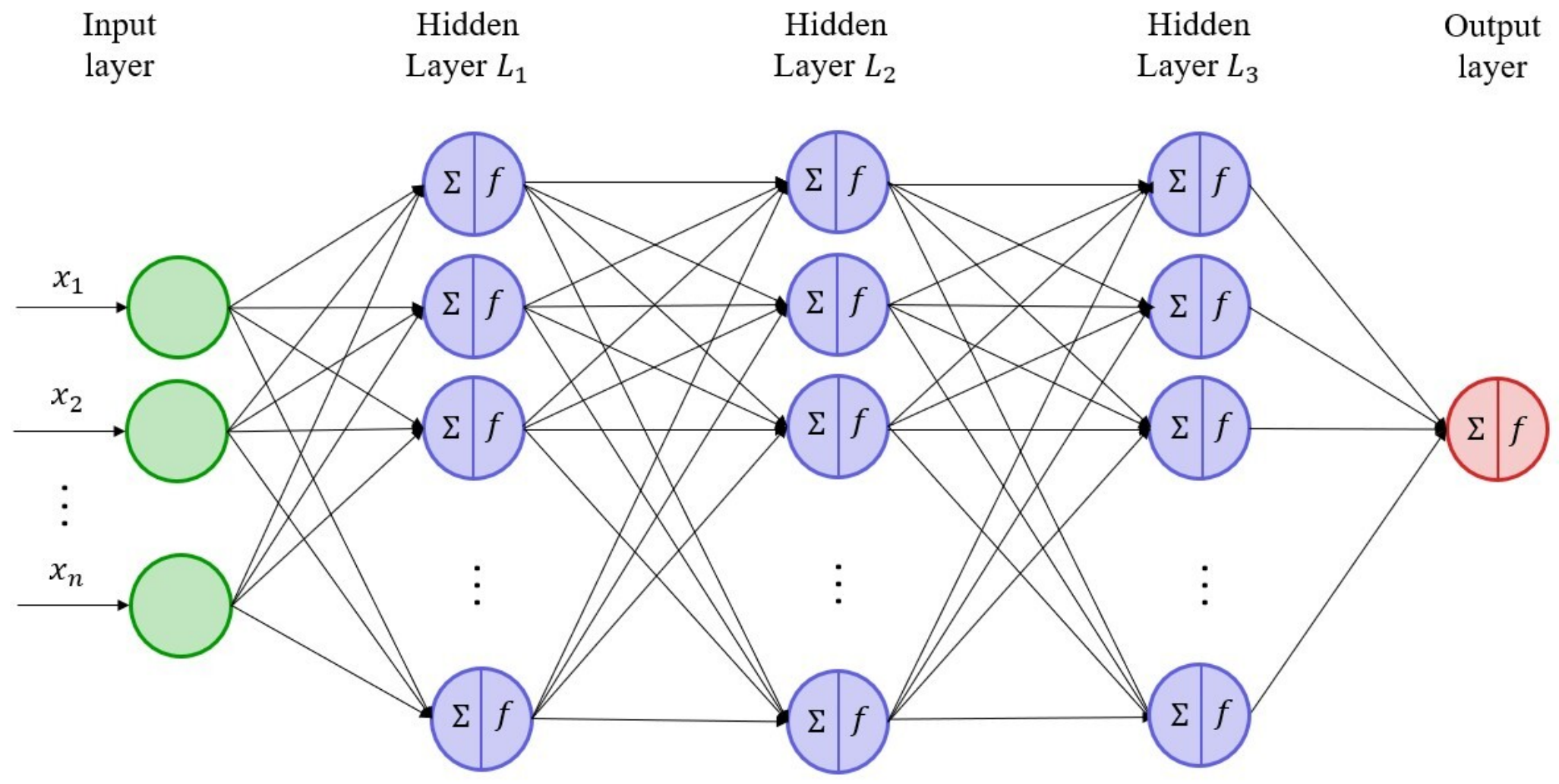}
        \caption{ANN structure of the calibration}
        \label{fig:fnn_structure}
    \end{figure}

    Using four training sets for $v_C^{Duan}$, $v_P^{Duan}$, $v_C^{CTS}$, and $v_P^{CTS}$, we train four multi-layer ANNs in this section. Each ANN consisted of three hidden layers with twenty nodes in each layer. The output is a single value, and the activation function of hidden layer nodes is the simple sigmoid function\footnote{$f(x) = \frac{1}{1+e^{-x}}$}, while the output activation function is the linear function\footnote{$f(x) = x$}. The ANNs for $v_C^{Duan}$ and $v_P^{Duan}$ have seven input nodes, while the ANNs for $v_C^{CTS}$, and $v_P^{CTS}$ have ten input nodes. The architecture is depicted in Fig. \ref{fig:fnn_structure}. 
    We denote the four ANNs corresponding to $v_C^{Duan}$, $v_P^{Duan}$, $v_C^{CTS}$, and $v_P^{CTS}$ as 
    \begin{align}
        &F_C^{Duan}(m, \tau; \kappa, \psi, \gamma, \theta, \sigma_0),
        \\
        &F_P^{Duan}(m, \tau; \kappa, \psi, \gamma, \theta, \sigma_0),
        \\
        &F_C^{CTS}(m, \tau; \kappa, \psi, \gamma, \theta, \sigma_0, \alpha, \lambda_+, \lambda_-),
        \\\text{and }
        &F_P^{CTS}(m, \tau; \kappa, \psi, \gamma, \theta, \sigma_0, \alpha, \lambda_+, \lambda_-).
    \end{align}
    Additionally, we set the parameters $\Theta_{Duan} = (\kappa$, $\psi$, $\gamma$, $\theta$, $\sigma_0$) of Duan's GARCH model and $\Theta_{CTS} = (\kappa$, $\psi$, $\gamma$, $\theta$, $\sigma_0$, $\alpha$, $\lambda_+$, $\lambda_-$) of CTS-GARCH model. Then, we have 
    \begin{align}
        \text{Call}_{ANN}^{Duan}(S_0, K, \tau, r) &= S_0 \exp\left(F_C^{Duan}\left(\frac{Ke^{-r\tau}}{S_0}, \tau; \Theta_{Duan}\right)\right)\\
        \text{Put}_{ANN}^{Duan}(S_0, K, \tau, r) &= S_0 \exp\left(F_P^{Duan}\left(\frac{Ke^{-r\tau}}{S_0}, \tau; \Theta_{Duan}\right)\right)\\
        \text{Call}_{ANN}^{CTS}(S_0, K, \tau, r) &= S_0 \exp\left(F_C^{CTS}\left(\frac{Ke^{-r\tau}}{S_0}, \tau; \Theta_{CTS}\right)\right)\\
        \text{Put}_{ANN}^{CTS}(S_0, K, \tau, r) &= S_0 \exp\left(F_P^{CTS}\left(\frac{Ke^{-r\tau}}{S_0}, \tau; \Theta_{CTS}\right)\right).
    \end{align}
    Subsequently, we obtain 
    \begin{align}
        \textup{Call}^{Duan}(S_0, K, \tau, r) &\approx \textup{Call}^{Duan}_{ANN}(S_0, K, \tau, r), \\
        \text{ and } 
        \textup{Put}^{Duan}(S_0, K, \tau, r) &\approx \textup{Put}^{Duan}_{ANN}(S_0, K,\tau, r),
    \end{align}
    under Duan's GARCH model, and    
    \begin{align}
        \textup{Call}^{CTS}(S_0, K, \tau, r) &\approx \textup{Call}^{CTS}_{ANN}(S_0, K, \tau, r), \\
        \text{ and }
        \textup{Put}^{CTS}(S_0, K, \tau, r) &\approx \textup{Put}^{CTS}_{ANN}(S_0, K,\tau, r),
    \end{align}
    under CTS-GARCH model.
    
    We train the four ANNs using functions using the Deep Learning toolbox in Matlab$^{TM}$. The default training algorithm for a function fitting network is Levenberg-Marquardt, as stated in the documentation\footnote{https://www.mathworks.com/help/deeplearning/ref/fitnet.html}. The mean squared error (MSE) is presented as the performance measure of the network. 
    
\begin{figure}
        \centering
        \includegraphics[width = 7cm]{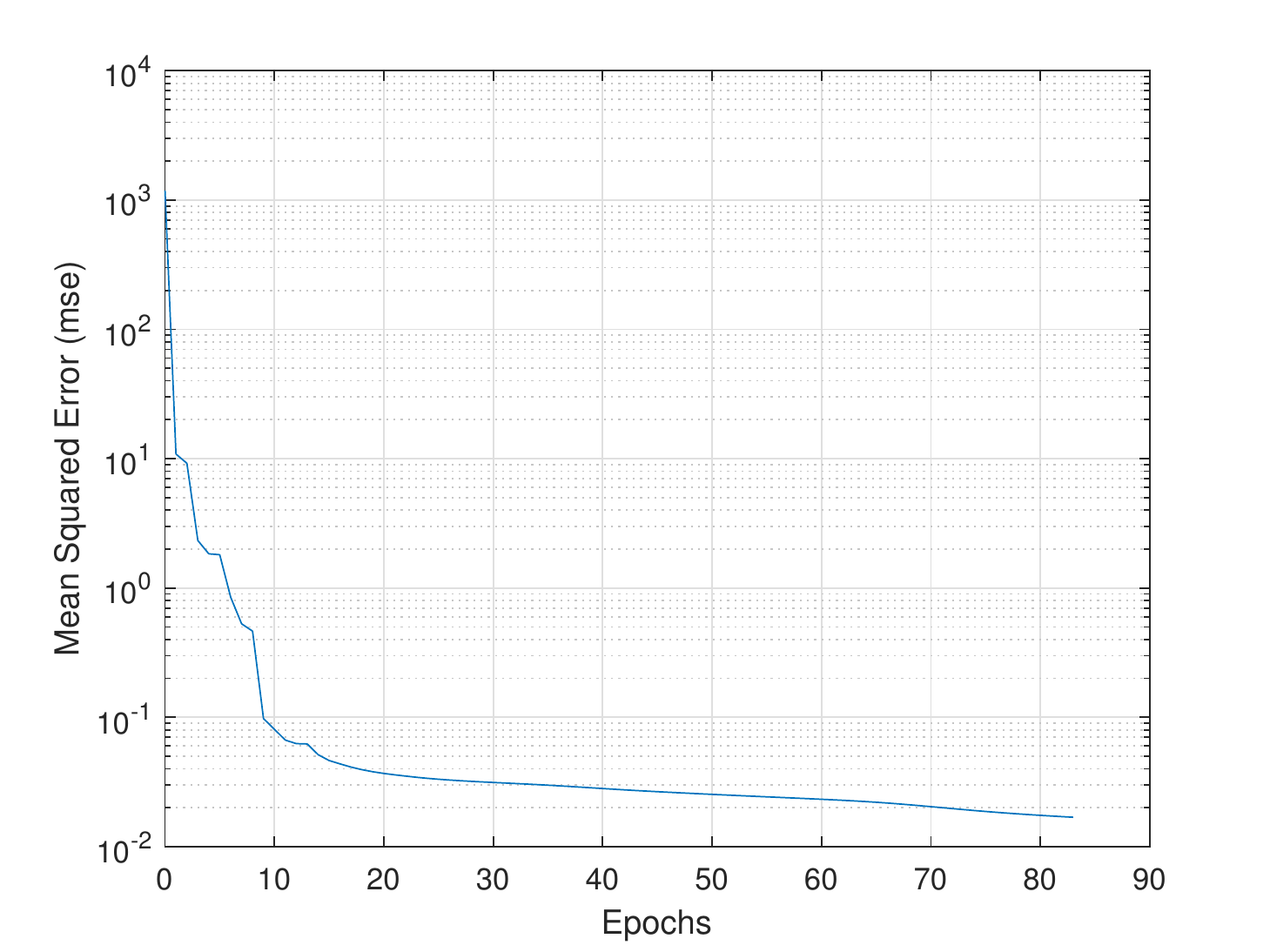}
        \includegraphics[width = 7cm]{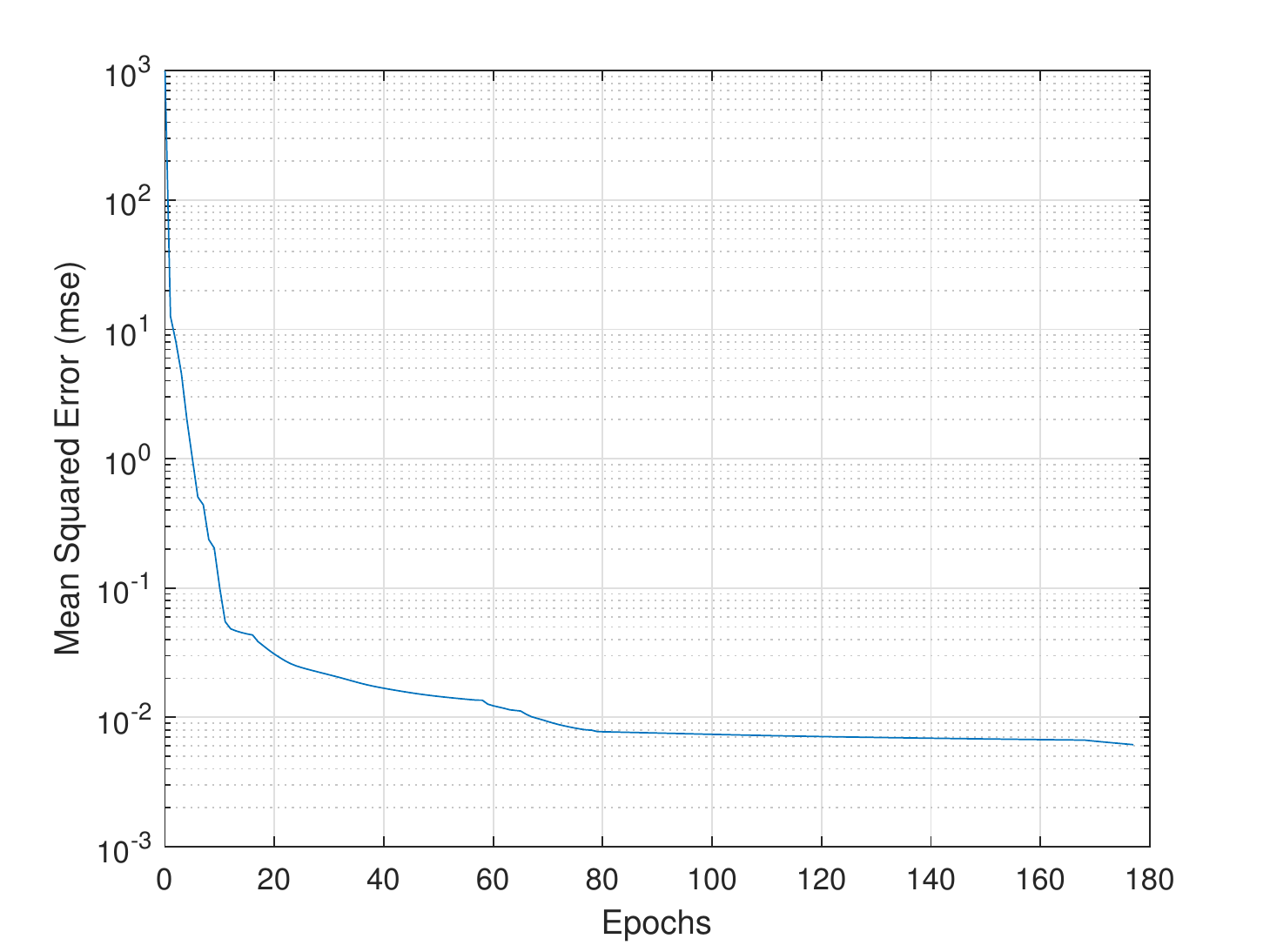}
        \includegraphics[width = 7cm]{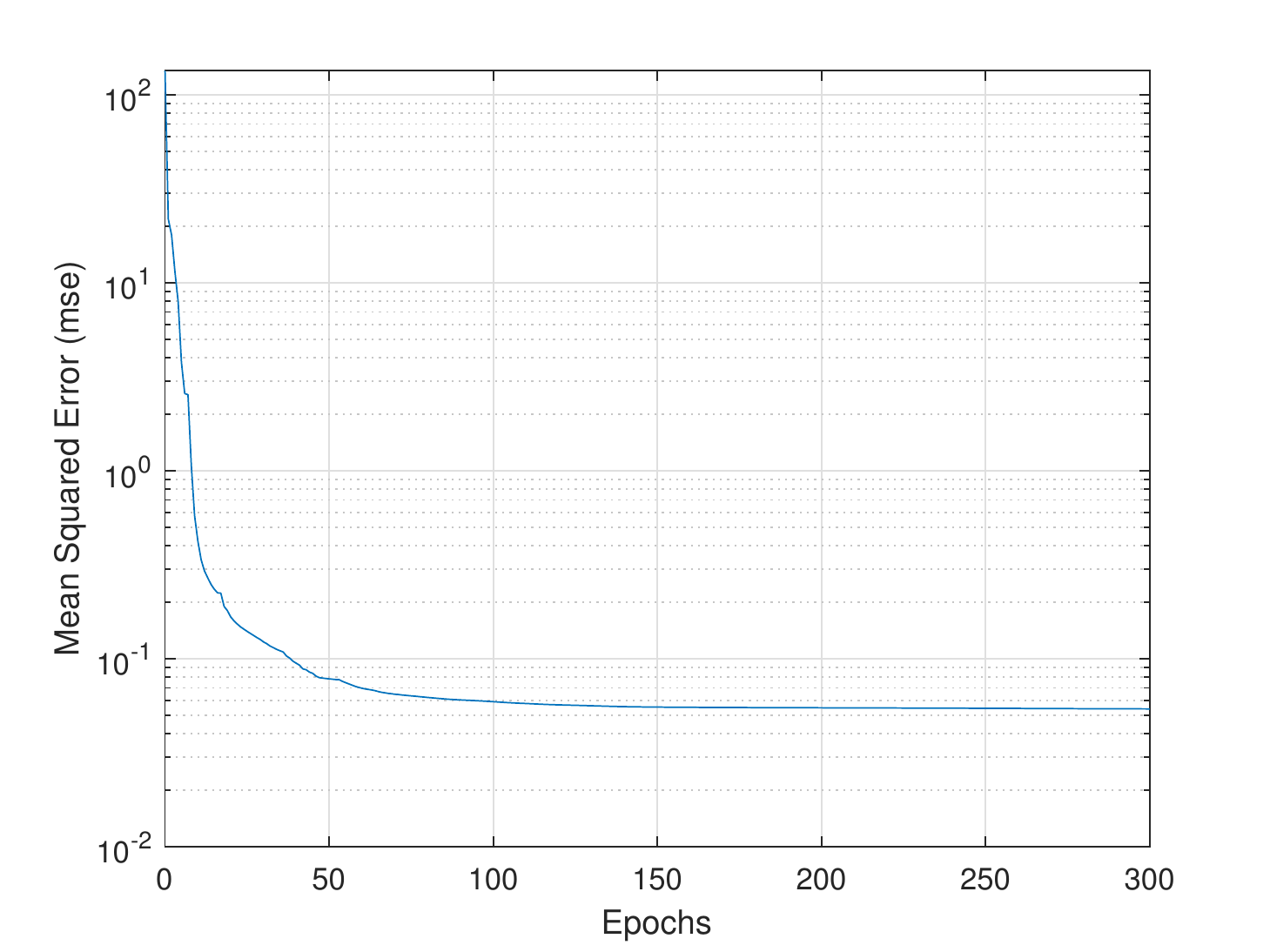}
        \includegraphics[width = 7cm]{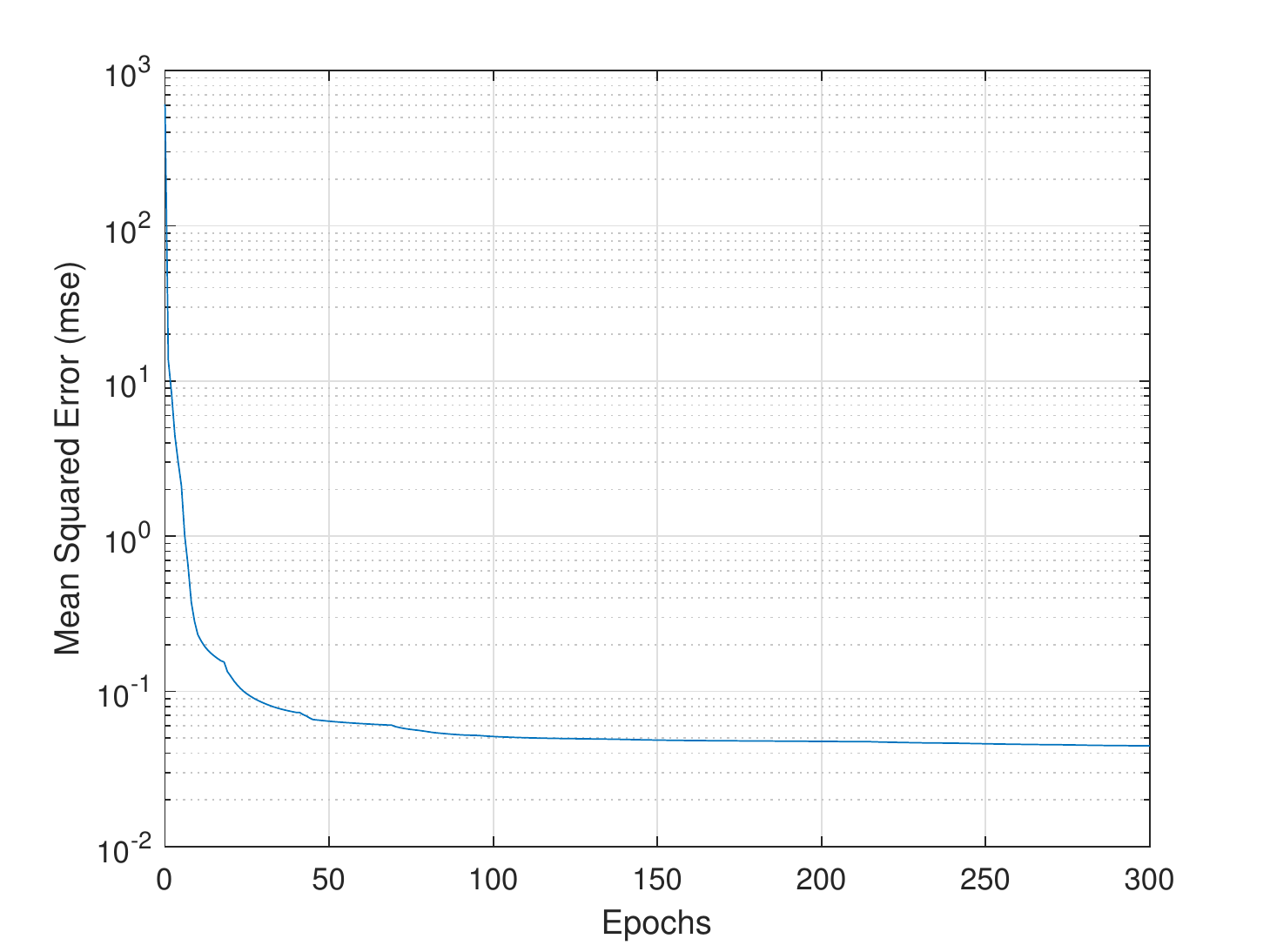}
        \caption{\label{fig:Epoch and mse} Number of epochs and mean squared error. Top left and right are the MSE of call and put prices for Duan's GARCH model. Bottom left and right are the MSE of the CTS-GARCH model.}
    \end{figure}

    Fig. \ref{fig:Epoch and mse} exhibits MSE for the epoch during the training of the four ANNs: $F_C^{Duan}$, $F_P^{Duan}$, $F_C^{CTS}$, and $F_P^{CTS}$. The minimum values of MSEs for $F_C^{Duan}$ and $F_P^{Duan}$ are  0.0169 and 0.0061 at epochs 84 and 178, respectively. For $F_C^{CTS}$ and $F_P^{CTS}$, the training was limited to 300 epochs, with minimum MSEs being 0.0541 and 0.0447 at epoch 300, respectively.

    It is also noteworthy that the ANNs in this study are not intended for forecasting, but rather for finding better analytic approximations. Therefore, we do not utilize a validation set.

\section{Calibration}
\label{Calibration}
    In this section, we discuss the calibration of parameters using two different methods: the ANN method and the MCS method, with the MCS method serving as a benchmark method. We calibrate the parameters $\Theta_{Duan}$ and $\Theta_{CTS}$ using call and put option data of the S\&P 500 index. For this investigation, we have selected 12 Wednesdays from the second week of each month, ranging from June 2021 to May 2022. The time to maturity of the option contracts varies between 7 and 90 days, and we exclude the options with zero bid prices or zero ask prices. Furthermore, we only calibrate parameters for out-of-the-money (OTM) options for both calls and puts. 
    
    We first specify ANNs for OTM options as
    \begin{align}
        F_{OTM}^{Duan}(m,\tau;\Theta_{Duan}) = F_P^{Duan}(m,\tau;\Theta_{Duan}) \cdot 1_{m<1} + F_C^{Duan}(m,\tau;\Theta_{Duan}) \cdot 1_{m\ge 1}
    \end{align}
    and 
    \begin{align}
        F_{OTM}^{CTS}(m,\tau;\Theta_{CTS})  =  F_P^{CTS}(m,\tau;\Theta_{CTS}) \cdot 1_{m<1} + F_C^{CTS}(m,\tau;\Theta_{CTS}) \cdot 1_{m\ge 1}.
    \end{align}
    We calibrate parameters using the relative root mean square error (rel-RMSE) minimization method as follows:
    \begin{align}
        \nonumber
        \min_{\Theta_{Duan}} \vast(&\sum_{K_ne^{-rT_n}<S_0} \left(\frac{F_{OTM}^{Duan}\left(\frac{K_ne^{-rT_n}}{S_0},\frac{T_n}{250};\Theta_{Duan}\right)-\log\left(\frac{P_{market}(K_n, T_n)}{S_0}\right)}{\log\left(\frac{P_{market}(K_n, T_n)}{S_0}\right)}\right)^2\\
        \label{eq:minSqErr_Duan}
        &+\sum_{K_ne^{-rT_n}\ge S_0} \left(\frac{F_{OTM}^{Duan}\left(\frac{K_ne^{-rT_n}}{S_0},\frac{T_n}{250};\Theta_{Duan}\right)-\log\left(\frac{C_{market}(K_n, T_n)}{S_0}\right)}{\log\left(\frac{C_{market}(K_n, T_n)}{S_0}\right)}\right)^2
        \vast)^{\frac{1}{2}}
    \end{align}
    and
    \begin{align}
        \nonumber
        \min_{\Theta_{CTS}} \vast(&\sum_{K_ne^{-rT_n}<S_0} \left(\frac{F_{OTM}^{CTS}\left(\frac{K_ne^{-rT_n}}{S_0},\frac{T_n}{250};\Theta_{CTS}\right)-\log\left(\frac{P_{market}(K_n, T_n)}{S_0}\right)}{\log\left(\frac{P_{market}(K_n, T_n)}{S_0}\right)}\right)^2\\
        \label{eq:minSqErr_CTS}
        &+\sum_{K_ne^{-rT_n}\ge S_0} \left(\frac{F_{OTM}^{CTS}\left(\frac{K_ne^{-rT_n}}{S_0},\frac{T_n}{250};\Theta_{CTS}\right)-\log\left(\frac{C_{market}(K_n, T_n)}{S_0}\right)}{\log\left(\frac{C_{market}(K_n, T_n)}{S_0}\right)}\right)^2
        \vast)^{\frac{1}{2}},
    \end{align}
    where $S_0$ is the S\&P 500 index price of the given Wednesday, and $C_{market}(K_n,T_n)$ and $P_{market}(K_n,T_n)$ are mid-prices of observed bid and ask prices for the call and put options with strike price $K_n$ and time to maturity $T_n$\footnote{To solve the nonlinear optimization problems, we used the function \texttt{lsqcurvefit()} in Matlab${}^{TM}$ with Trust-Region-Reflective Least Squares algorithm. See \cite{ColemanLi:1994,ColemanLi:1996} and \url{https://www.mathworks.com/help/optim/ug/lsqcurvefit.html} for the details.}.
    
    As a benchmark, we use the MCS method to calibrate the same parameters. Similar to the ANN cases, we define
    \begin{align}
        v_{OTM}^{Duan}(m,\tau;\Theta_{Duan}) = v_C^{Duan}(m,\tau;\Theta_{Duan}) \cdot 1_{m\ge 1} + v_P^{Duan}(m,\tau;\Theta_{Duan}) \cdot 1_{m<1}
    \end{align}
    and 
    \begin{align}
        v_{OTM}^{CTS}(m,\tau;\Theta_{CTS})  = v_C^{CTS}(m,\tau;\Theta_{CTS}) \cdot 1_{m\ge 1} + v_P^{CTS}(m,\tau;\Theta_{CTS}) \cdot 1_{m<1}.
    \end{align}
    We calibrate parameters using the MCS method as described in \eqref{eq:minSqErr_Duan} and \eqref{eq:minSqErr_CTS} by replacing $F_{OTM}^{Duan}$ and $F_{OTM}^{CTS}$ with $v_{OTM}^{Duan}$ and $v_{OTM}^{CTS}$, respectively. More details of the parameter calibration for the option pricing with the CTS-GARCH model are presented in \cite{Kim_et_al:2010:JBF}, \cite{KimDanlingStoyan:2019}, and \cite{KimRhoDouady:2022}.
    
    The calibrated parameters for Duan's GARCH model are provided in Table \ref{tbl:MCS Cal Duan} and Table \ref{tbl:ANN Cal Duan}, while those for the CTS-GARCH model are provided in Table \ref{tbl:MCS Cal CTS} and Table \ref{tbl:ANN Cal CTS}. Table \ref{tbl:ANN Cal Duan} and \ref{tbl:ANN Cal CTS} present the results obtained using the ANN method, while Table \ref{tbl:MCS Cal Duan} and Table \ref{tbl:MCS Cal CTS} present the results obtained using the MCS method. The rel-RMSE values are presented for performance analysis. Table \ref{tbl:RelRMSErr} collects all rel-RMSEs for the two models (Duan's GARCH and CTS-GARCH) and both option pricing methods (MCS and ANN). 
    
    According to the table, we observe that:
    \begin{itemize}
        \item CTS-GARCH model with the ANN method shows the smallest rel-RMSE  and therefore performed the best, with the exception of four cases on 11/10/2022, 12/8/2021, 3/9/2022, and 5/10/2022.
        \item Duan's GARCH model with the MCS method has the smallest rel-RMSE for the cases on 3/8/2022 and 5/10/2022.
        \item Duan's GARCH model with the ANN method has the smallest rel-RMSE for the cases on 11/10/2022 and 12/8/2022.
        \item CTS-GARCH model with the MCS method has the largest rel-RMSE values in this investigation. 
    \end{itemize}
    Compared to Duan's GARCH model, CTS-GARCH model is more flexible as it has three more parameters. However, the complexity of the CTS-GARCH model makes model calibration with the MCS method inefficient. In this regard, the ANN offers an efficient alternative for model calibration, enabling the use of the CTS-GARCH model in practical applications.
       
\begin{figure}
    \centering
    \includegraphics[width = 7cm]{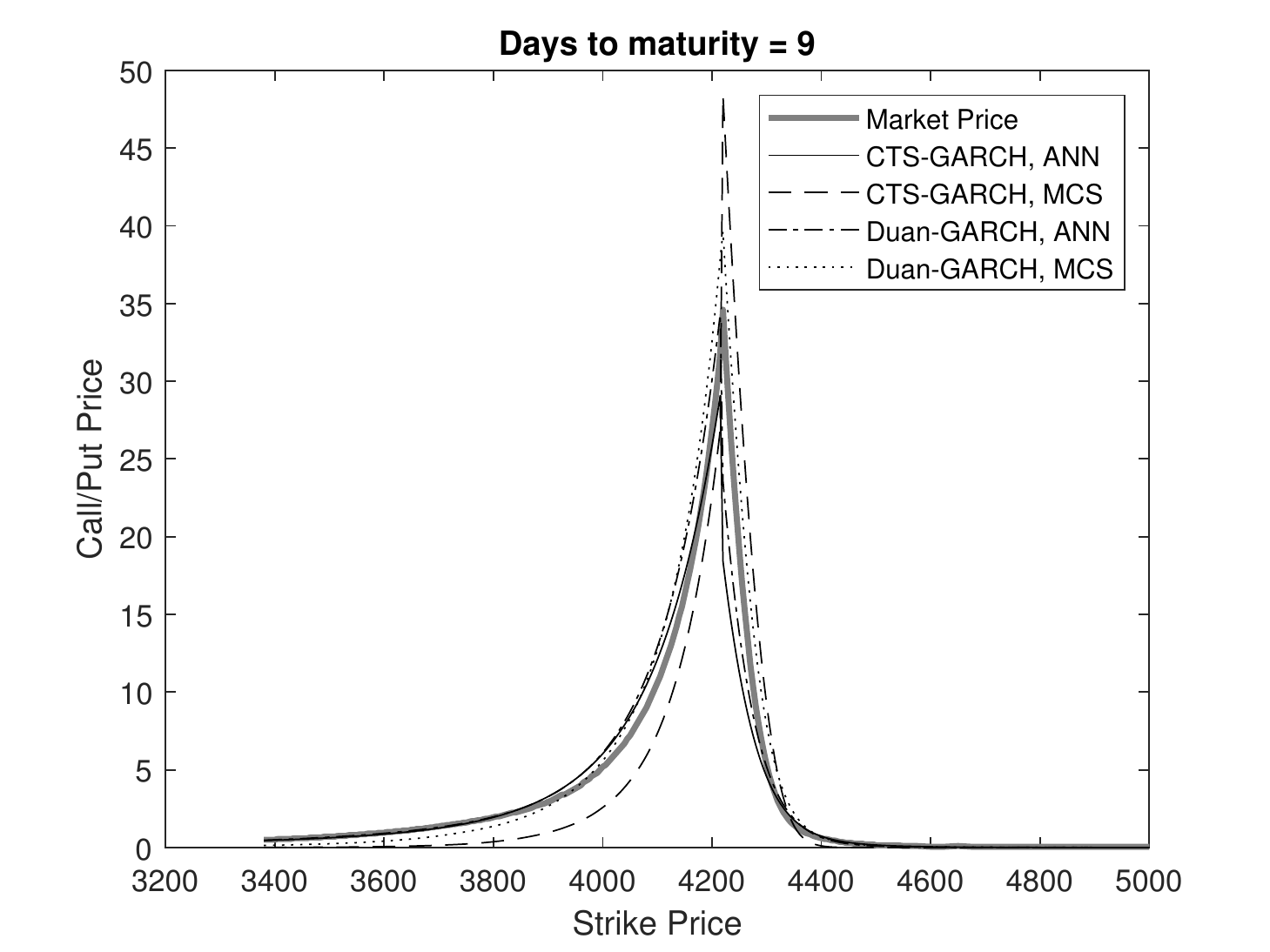}
    \includegraphics[width = 7cm]{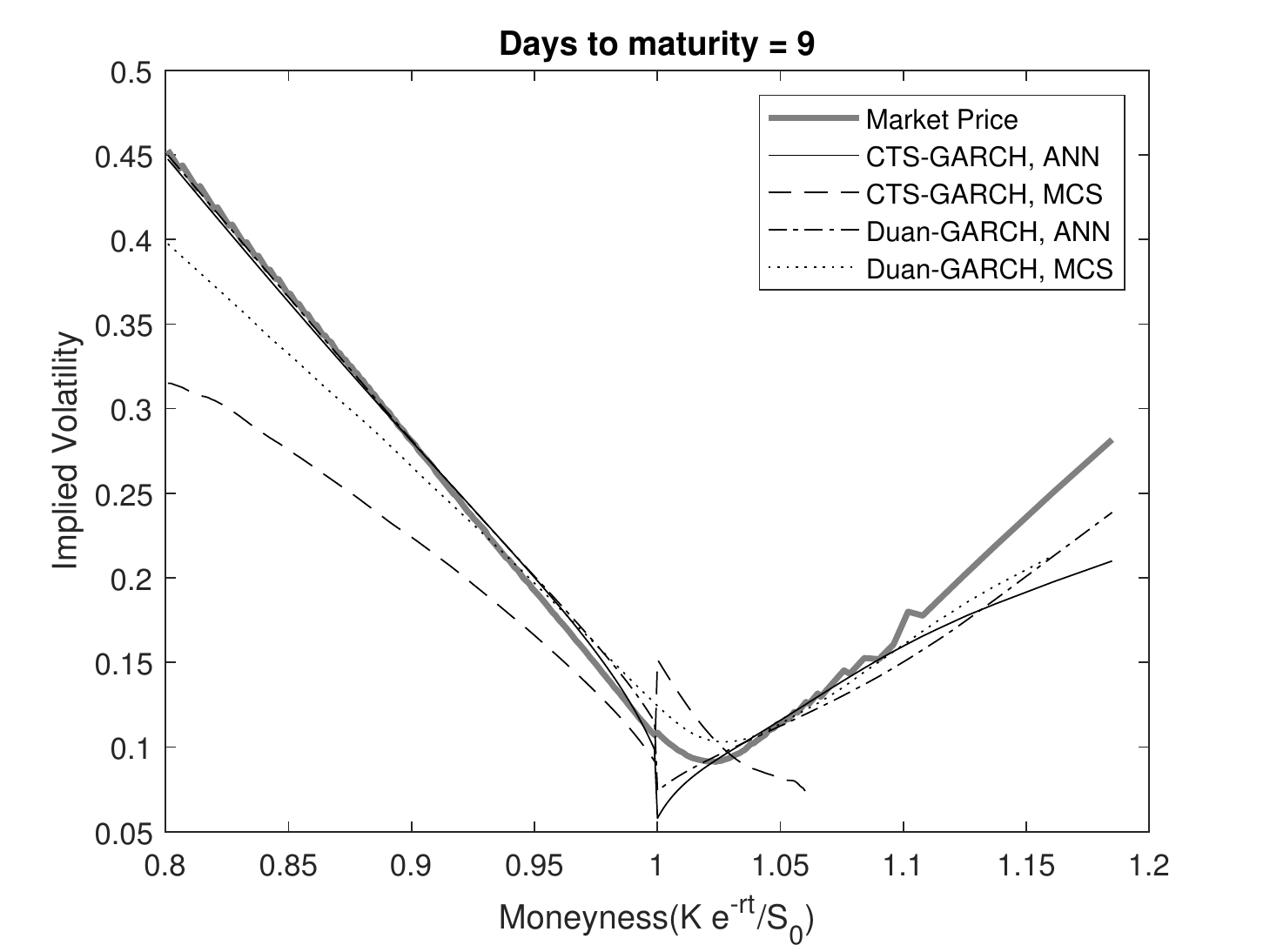}\\
    \includegraphics[width = 7cm]{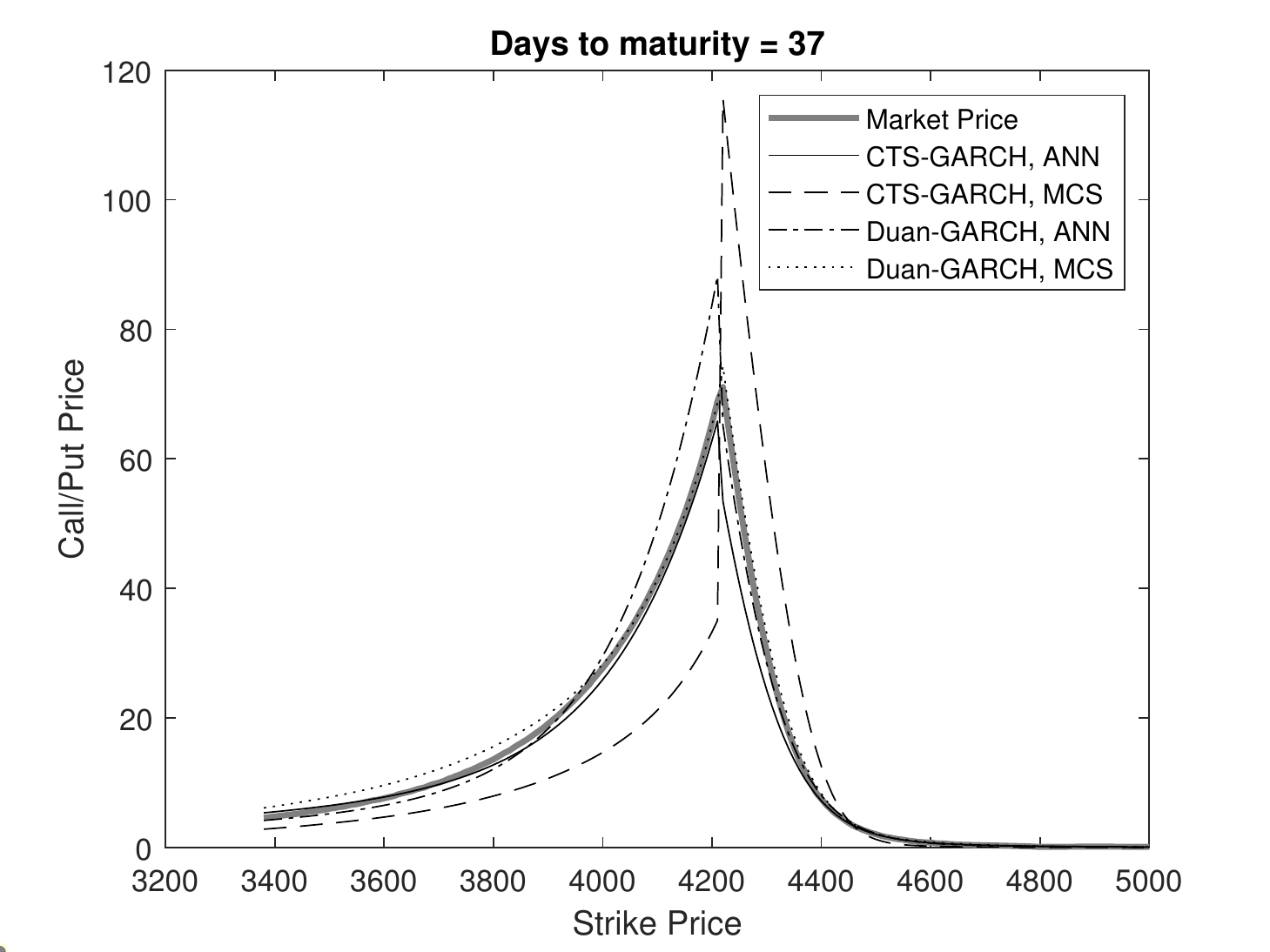}
    \includegraphics[width = 7cm]{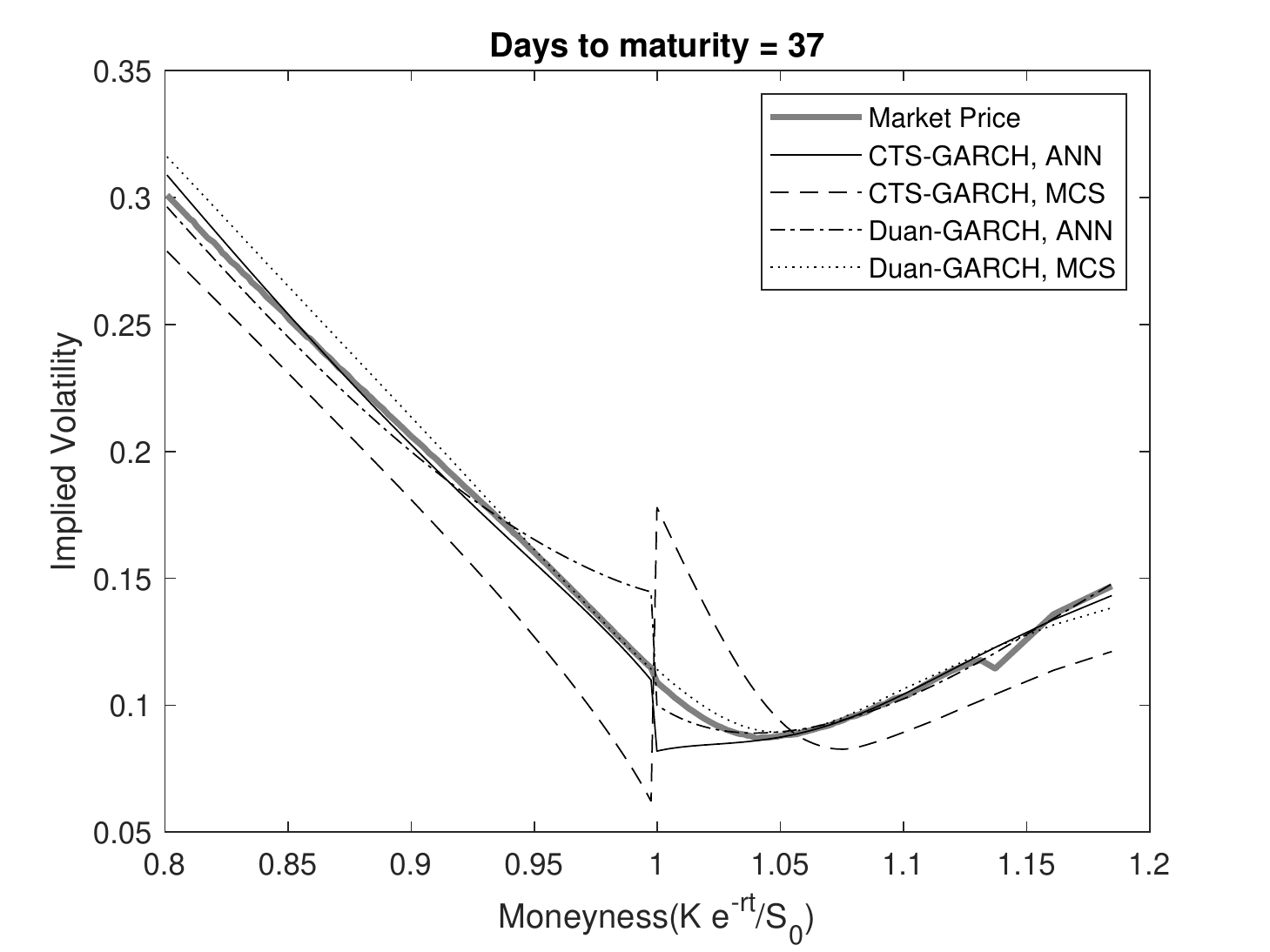}\\
    \includegraphics[width = 7cm]{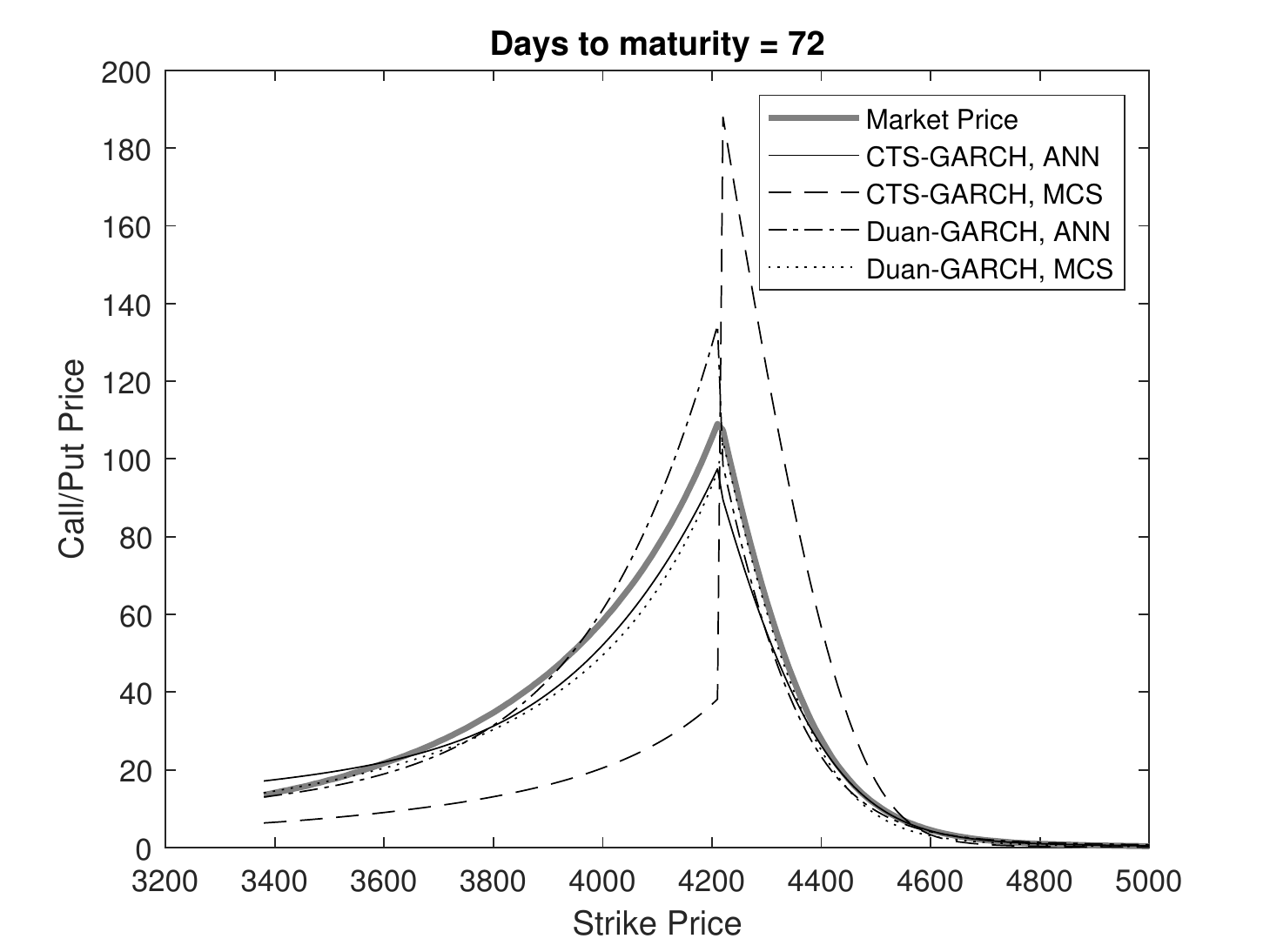}
    \includegraphics[width = 7cm]{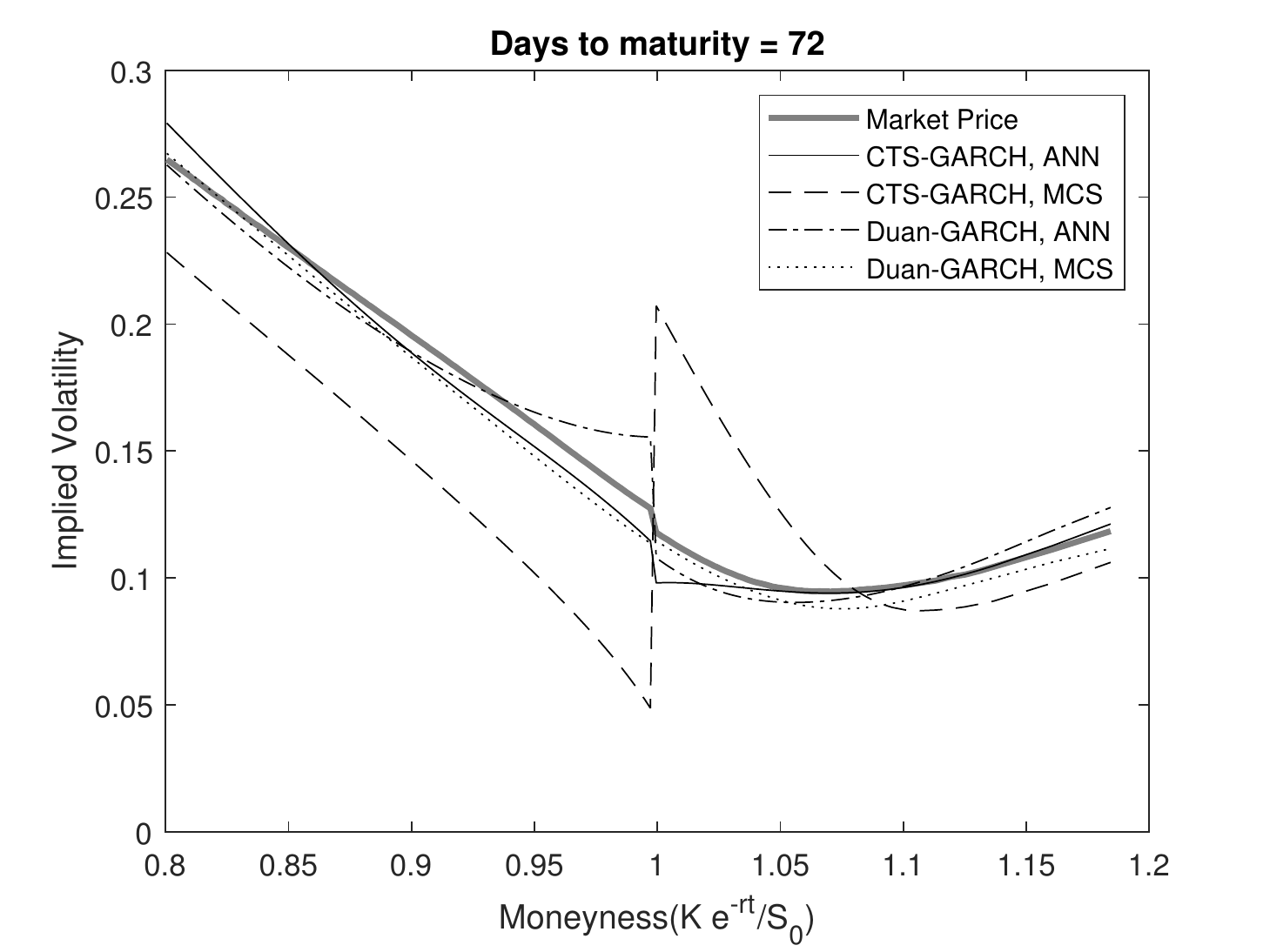}
    \caption{\label{fig:CallPutPriceImpvol} Calibrated call and put option prices and implied volatility curses at 6/9/2021}
    \end{figure}

    In Fig. \ref{fig:CallPutPriceImpvol}, the left column plots present the market prices and calibrated model prices of the Duan's GARCH model and CTS-GARCH model for OTM calls and puts on 6/9/2021. The right column plots exhibit the implied volatility curves for the market prices and Duan's GARCH and CTS-GARCH models with respect to the MCS method and ANN method. The first row plots of Fig. \ref{fig:CallPutPriceImpvol} show option prices and implied volatility for 9 days to maturity, the second row plots show 37 days to maturity, and the third row plots show 72 days to maturity. The curves suggest that the implied volatility curve of CTS-GARCH model prices with the ANN method is the closest to the market implied volatility curve compared to the other methods investigated in this study.
    
    One of the advantages of using the ANN method is the ability to easily obtain the Greeks of options, which is not almost possible with the MCS method. For instance, Table \ref{tbl:DeltaGamma} exhibits Delta ($\Delta$), Gamma ($\Gamma$), Theta ($\Theta$) and Rho ($\rho$) of at-the-money (ATM) calls and puts for the parameters calibrated on 6/9/2021 with maturities of 7, 37, and 72 days, respectively\footnote{Delta $\left(\Delta=\frac{\partial V}{\partial S}\right)$: the sensitivity of an option’s price($V$) changes relative to the underlying price($S$) changes.\\ 
    \indent Gamma $\left(\Gamma=\frac{\partial^2 V}{\partial S^2}\right)$: the sensitivity of the Delta's changes relative to the underlying price changes.\\\indent Theta $\left(\Theta=\frac{\partial V}{\partial \tau}\right)$: the sensitivity of the option price relative to the option’s time to maturity($\tau$). \\\indent Rho $\left(\rho=\frac{\partial V}{\partial r}\right)$: the sensitivity of the option price relative to the risk-free rate of return ($r$).\\\indent Vega $\left(\nu=\frac{\partial V}{\partial \sigma}\right)$: the sensitivity of an option’s price changes relative to the underlying asset's volatility.}. The Vega ($\nu$) is not considered since the volatility is not constant but stochastic in this investigation. We calculate Greeks using the finance difference method for each variable. 
    
    It is widely acknowledged that the calculation time of the ANN method is remarkably faster than the MCS method. Table \ref{tbl:timming} presents the calculation time for OTM calls and puts using the MCS and ANN methods on Duan's GARCH and CTS-GARCH models, respectively. For example, we observed 558 prices of calls and puts on 6/9/2021. The MCS and ANN methods respectively take 0.13 and 0.085 seconds to calculate 558 calls and puts on Duan's GARCH model, and 0.3286 and 0.0187 seconds on the CTS-GARCH model. The number of call and put price observations (Obs.) varies each day and is shown in the Obs. column in Table \ref{tbl:timming}. The calculation times using the ANN method of Duan's GARCH model are about 9 times faster than the MCS method. Similarly, in the CTS-GARCH model case, the MCS method takes approximately 20 times longer than the ANN method.
    
    \begin{table}
    \centering
    \begin{footnotesize}
    \begin{tabular}{c|ccccc|c}
    \hline
    Date & $\theta$ & $\kappa$ & $\xi$ & $\zeta$ & $\sigma_0$ & rel-RMSE\\
    \hline
    6/9/2021	& $	0.9293	$ & $	1.68\cdot 10^{-6}	$ & $	0.2778	$ & $	0.5000	$ & $	0.0085	$ & $	0.3048	$ \\
    7/7/2021	& $	0.8984	$ & $	1.21\cdot 10^{-6}	$ & $	0.2931	$ & $	0.5000	$ & $	0.0068	$ & $	0.3585	$ \\
    8/11/2021	& $	0.9513	$ & $	1.05\cdot 10^{-6}	$ & $	0.2847	$ & $	0.5000	$ & $	0.0057	$ & $	0.3927	$ \\
    9/8/2021	& $	1.0877	$ & $	9.88\cdot 10^{-7}	$ & $	0.2492	$ & $	0.5000	$ & $	0.0065	$ & $	0.4349	$ \\
    10/6/2021	& $	0.9826	$ & $	2.70\cdot 10^{-6}	$ & $	0.2584	$ & $	0.5000	$ & $	0.0094	$ & $	0.2837	$ \\
    11/10/2021	& $	0.9905	$ & $	1.65\cdot 10^{-6}	$ & $	0.2681	$ & $	0.5000	$ & $	0.0075	$ & $	0.2952	$ \\
    12/8/2021	& $	0.8833	$ & $	2.27\cdot 10^{-6}	$ & $	0.2943	$ & $	0.5000	$ & $	0.0081	$ & $	0.2763	$ \\
    1/12/2022	& $	0.8447	$ & $	2.13\cdot 10^{-6}	$ & $	0.3000	$ & $	0.5000	$ & $	0.0073	$ & $	0.2799	$ \\
    2/9/2022	& $	0.8745	$ & $	2.94\cdot 10^{-6}	$ & $	0.2869	$ & $	0.5000	$ & $	0.0091	$ & $	0.2305	$ \\
    3/9/2022	& $	1.0237	$ & $	5.27\cdot 10^{-6}	$ & $	0.2464	$ & $	0.5000	$ & $	0.0158	$ & $	0.1881	$ \\
    4/6/2022	& $	1.0867	$ & $	1.47\cdot 10^{-6}	$ & $	0.2434	$ & $	0.5000	$ & $	0.0092	$ & $	0.3980	$ \\
    5/10/2022	& $	1.0220	$ & $	3.97\cdot 10^{-6}	$ & $	0.1918	$ & $	0.5971	$ & $	0.0176	$ & $	0.2311	$ \\
    \hline
    \end{tabular}
    \end{footnotesize}
    \caption{\label{tbl:MCS Cal Duan}Calibration of Duan's GARCH model parameters using the MCS method}
    \end{table}
    
    \begin{table}
    \centering
    \begin{footnotesize}
    \begin{tabular}{c|ccccc|c}
    \hline
    Date & $\theta$ & $\kappa$ & $\xi$ & $\zeta$ & $\sigma_0$ & rel-RMSE\\
    \hline
    6/9/2021	& $	1.1103	$ & $	4.64\cdot 10^{-7}	$ & $	0.2258	$ & $	0.5395	$ & $	0.0064	$ & $	0.1972	$ \\
    7/7/2021	& $	1.0196	$ & $	6.32\cdot 10^{-7}	$ & $	0.2354	$ & $	0.5621	$ & $	0.0049	$ & $	0.2793	$ \\
    8/11/2021	& $	1.1729	$ & $	1.74\cdot 10^{-7}	$ & $	0.1994	$ & $	0.5737	$ & $	0.0056	$ & $	0.3383	$ \\
    9/8/2021	& $	1.1810	$ & $	2.89\cdot 10^{-7}	$ & $	0.1927	$ & $	0.5783	$ & $	0.0064	$ & $	0.4107	$ \\
    10/6/2021	& $	0.9015	$ & $	8.75\cdot 10^{-7}	$ & $	0.2115	$ & $	0.6441	$ & $	0.0074	$ & $	0.2826	$ \\
    11/10/2021	& $	1.1827	$ & $	4.99\cdot 10^{-7}	$ & $	0.1968	$ & $	0.5700	$ & $	0.0066	$ & $	0.2634	$ \\
    12/8/2021	& $	1.1276	$ & $	7.31\cdot 10^{-7}	$ & $	0.2238	$ & $	0.5424	$ & $	0.0060	$ & $	0.2078	$ \\
    1/12/2022	& $	0.6377	$ & $	4.46\cdot 10^{-7}	$ & $	0.3000	$ & $	0.6252	$ & $	0.0047	$ & $	0.3632	$ \\
    2/9/2022	& $	0.8367	$ & $	6.29\cdot 10^{-7}	$ & $	0.2081	$ & $	0.6739	$ & $	0.0076	$ & $	0.2626	$ \\
    3/9/2022	& $	0.9301	$ & $	1.64\cdot 10^{-6}	$ & $	0.1905	$ & $	0.6666	$ & $	0.0121	$ & $	0.2297	$ \\
    4/6/2022	& $	1.0649	$ & $	5.37\cdot 10^{-7}	$ & $	0.2088	$ & $	0.5990	$ & $	0.0070	$ & $	0.3616	$ \\
    5/10/2022	& $	1.0130	$ & $	1.67\cdot 10^{-6}	$ & $	0.1506	$ & $	0.7046	$ & $	0.0141	$ & $	0.2417	$ \\
    \hline
    \end{tabular}
    \end{footnotesize}
    \caption{\label{tbl:ANN Cal Duan}Calibration of Duan's GARCH model parameters using the ANN method}
    \end{table}
    
    \begin{table}
    \centering
    \begin{footnotesize}
    \begin{tabular}{c|cccccccc|c}
    \hline
    Date & $\theta$ & $\kappa$ & $\xi$ & $\zeta$ & $\sigma_0$ & $\alpha$ & $\lambda_+$ & $\lambda_-$ & rel-RMSE\\
    \hline
    6/9/2021	& $	0.9386	$ & $	1.69\cdot 10^{-6}	$ & $	0.2805	$ & $	0.5050	$ & $	0.0086	$ & $	1.9998	$ & $	14.0138	$ & $	10.0760	$ & $	0.8819	$ \\
    7/7/2021	& $	0.9074	$ & $	1.23\cdot 10^{-6}	$ & $	0.2960	$ & $	0.5050	$ & $	0.0068	$ & $	1.9921	$ & $	11.2482	$ & $	11.9459	$ & $	0.9014	$ \\
    8/11/2021	& $	0.9608	$ & $	1.06\cdot 10^{-6}	$ & $	0.2876	$ & $	0.5050	$ & $	0.0057	$ & $	1.9998	$ & $	11.9686	$ & $	6.3593	$ & $	0.9211	$ \\
    9/8/2021	& $	1.0986	$ & $	9.97\cdot 10^{-7}	$ & $	0.2517	$ & $	0.5050	$ & $	0.0066	$ & $	1.9543	$ & $	13.9703	$ & $	87.5816	$ & $	0.9113	$ \\
    10/6/2021	& $	0.9923	$ & $	2.72\cdot 10^{-6}	$ & $	0.2609	$ & $	0.5050	$ & $	0.0095	$ & $	1.9833	$ & $	51.2546	$ & $	9.3704	$ & $	0.8716	$ \\
    11/10/2021	& $	1.0004	$ & $	1.67\cdot 10^{-6}	$ & $	0.2708	$ & $	0.5050	$ & $	0.0076	$ & $	1.8757	$ & $	13.9376	$ & $	11.8176	$ & $	0.8938	$ \\
    12/8/2021	& $	0.8921	$ & $	2.30\cdot 10^{-6}	$ & $	0.2973	$ & $	0.5050	$ & $	0.0082	$ & $	1.9999	$ & $	6.0208	$ & $	24.1959	$ & $	0.8832	$ \\
    1/12/2022	& $	0.8531	$ & $	2.15\cdot 10^{-6}	$ & $	0.3030	$ & $	0.5050	$ & $	0.0073	$ & $	1.7526	$ & $	5.2020	$ & $	27.2385	$ & $	0.8971	$ \\
    2/9/2022	& $	0.8832	$ & $	2.97\cdot 10^{-6}	$ & $	0.2898	$ & $	0.5050	$ & $	0.0092	$ & $	1.9999	$ & $	11.2642	$ & $	9.8365	$ & $	0.8822	$ \\
    3/9/2022	& $	1.0340	$ & $	5.32\cdot 10^{-6}	$ & $	0.2488	$ & $	0.5050	$ & $	0.0160	$ & $	0.8949	$ & $	8.5344	$ & $	28.3573	$ & $	0.8142	$ \\
    4/6/2022	& $	1.0976	$ & $	1.48\cdot 10^{-6}	$ & $	0.2458	$ & $	0.5050	$ & $	0.0093	$ & $	1.9999	$ & $	2.4136	$ & $	0.7439	$ & $	0.8864	$ \\
    5/10/2022	& $	1.0323	$ & $	4.01\cdot 10^{-6}	$ & $	0.1937	$ & $	0.6031	$ & $	0.0178	$ & $	1.9984	$ & $	14.8261	$ & $	15.2165	$ & $	0.7809	$ \\
    
    \hline
    \end{tabular}
    \end{footnotesize}
    \caption{\label{tbl:MCS Cal CTS}Calibration of CTS-GARCH model parameters using the MCS method}
    \end{table}
    
    \begin{table}
    \centering
    \begin{footnotesize}
    \begin{tabular}{c|cccccccc|c}
    \hline
    Date & $\theta$ & $\kappa$ & $\xi$ & $\zeta$ & $\sigma_0$ & $\alpha$ & $\lambda_+$ & $\lambda_-$ & rel-RMSE\\
    \hline
    6/9/2021	& $	0.5526	$ & $	1.56\cdot 10^{-8}	$ & $	0.0954	$ & $	0.7906	$ & $	0.0046	$ & $	1.4202	$ & $	0.1019	$ & $	0.5492	$ & $	0.1662	$ \\
    7/7/2021	& $	2.5598	$ & $	4.65\cdot 10^{-6}	$ & $	0.0631	$ & $	0.5124	$ & $	0.0065	$ & $	1.0542	$ & $	13.6212	$ & $	0.1155	$ & $	0.2372	$ \\
    8/11/2021	& $	1.2173	$ & $	1.43\cdot 10^{-8}	$ & $	0.1960	$ & $	0.6349	$ & $	0.0076	$ & $	0.7306	$ & $	46.0982	$ & $	0.1629	$ & $	0.2936	$ \\
    9/8/2021	& $	1.1724	$ & $	6.67\cdot 10^{-10}	$ & $	0.1582	$ & $	0.7666	$ & $	0.0068	$ & $	0.8133	$ & $	3.9428	$ & $	0.1019	$ & $	0.2781	$ \\
    10/6/2021	& $	0.5119	$ & $	2.51\cdot 10^{-8}	$ & $	0.0014	$ & $	0.5016	$ & $	0.0163	$ & $	0.5610	$ & $	13.7536	$ & $	12.3532	$ & $	0.2504	$ \\
    11/10/2021	& $	1.9605	$ & $	4.93\cdot 10^{-8}	$ & $	0.1048	$ & $	0.5092	$ & $	0.0079	$ & $	0.6510	$ & $	1.4461	$ & $	4.1668	$ & $	0.2701	$ \\
    12/8/2021	& $	0.0281	$ & $	3.46\cdot 10^{-6}	$ & $	0.2965	$ & $	0.7125	$ & $	0.0091	$ & $	0.7452	$ & $	31.5555	$ & $	0.5439	$ & $	0.2474	$ \\
    1/12/2022	& $	0.6766	$ & $	2.27\cdot 10^{-6}	$ & $	0.2979	$ & $	0.5750	$ & $	0.0075	$ & $	0.5862	$ & $	20.2640	$ & $	3.5172	$ & $	0.2233	$ \\
    2/9/2022	& $	0.4874	$ & $	2.72\cdot 10^{-6}	$ & $	0.2624	$ & $	0.6696	$ & $	0.0100	$ & $	0.7577	$ & $	27.9998	$ & $	1.5475	$ & $	0.2152	$ \\
    3/9/2022	& $	0.5884	$ & $	2.91\cdot 10^{-6}	$ & $	0.1191	$ & $	0.8183	$ & $	0.0134	$ & $	0.6454	$ & $	21.7075	$ & $	4.0287	$ & $	0.2684	$ \\
    4/6/2022	& $	1.5418	$ & $	9.84\cdot 10^{-6}	$ & $	0.1040	$ & $	0.5271	$ & $	0.0078	$ & $	0.9238	$ & $	21.6161	$ & $	1.5811	$ & $	0.3571	$ \\
    5/10/2022	& $	0.8144	$ & $	3.15\cdot 10^{-8}	$ & $	0.1669	$ & $	0.7220	$ & $	0.0178	$ & $	0.7674	$ & $	28.8533	$ & $	4.4167	$ & $	0.2756	$ \\
    \hline
    \end{tabular}
    \end{footnotesize}
    \caption{\label{tbl:ANN Cal CTS}Calibration of CTS-GARCH model parameters using the ANN method}
    \end{table}
    
    \begin{table}
    \centering
    \begin{tabular}{cccccc}
    \hline
     & \multicolumn{2}{c}{Duan's GARCH Model} & & \multicolumn{2}{c}{CTS-GARCH Model} \\
     \cline{2-3}  \cline{5-6}
    Date & MCS & ANN & & MCS & ANN \\
    \hline
    6/9/2021	& $	0.3048	$ & $	0.1972	$ & & $	0.8819	$ & $\textbf{0.1662}	$ \\
    7/7/2021	& $	0.3585	$ & $	0.2793	$ & & $	0.9014	$ & $\textbf{0.2372}	$ \\
    8/11/2021	& $	0.3927	$ & $	0.3383	$ & & $	0.9211	$ & $	\textbf{0.2936}	$ \\
    9/8/2021	& $	0.4349	$ & $	0.4107	$ & & $	0.9113	$ & $\textbf{0.2781}	$ \\
    10/6/2021	& $	0.2837	$ & $	0.2826	$ & & $	0.8716	$ & $	\textbf{0.2504}	$ \\
    11/10/2021	& $	0.2952	$ & $\textbf{0.2634}	$ & & $	0.8938	$ & $	0.2701	$ \\
    12/8/2021	& $	0.2763	$ & $\textbf{0.2078}	$ & & $	0.8832	$ & $	0.2474	$ \\
    1/12/2022	& $	0.2799	$ & $	0.3632	$ & & $	0.8971	$ & $	\textbf{0.2233}	$ \\
    2/9/2022	& $	0.2305	$ & $	0.2626	$ & & $	0.8822	$ & $\textbf{0.2152}	$ \\
    3/9/2022	& $\textbf{0.1881}	$ & $0.2297	$ & & $	0.8142	$ & $	0.2684	$ \\
    4/6/2022	& $	0.3980	$ & $	0.3616	$ & & $	0.8864	$ & $\textbf{0.3571}	$ \\
    5/10/2022	& $\textbf{0.2311}	$ & $	0.2417	$ & & $	0.7809	$ & $	0.2756	$ \\
    \hline
    \end{tabular}
    \caption{\label{tbl:RelRMSErr} Relative RMSE values for model calibrations}
    \end{table}
    
    \begin{table}
    \centering
    \begin{footnotesize}
    \begin{tabular}{cccccccccc}
    \hline
    Call/Put & $S_0$ & $r$ & $K$ & $\tau$ & Model & $\Delta$ & $\Gamma$ & $\Theta$ & $\rho$\\
    \hline
    Call & $4219.55$ & $0.0020$ & $4220$ & $0.0360$ &  Duan's GARCH & $0.4109$ & $5.7074\cdot 10^{-3}$ & $382.59$ & $61.56$  \\ 
     & & & & & CTS-GARCH & $0.2933$ & $3.5068\cdot 10^{-3}$ & $274.16$ & $43.88$  \\ 
     & & & &$0.1480$ &  Duan's GARCH & $0.6089$ & $3.2794\cdot 10^{-3}$ & $311.14$ & $370.61$  \\ 
     & & & & & CTS-GARCH & $0.4744$ & $2.4125\cdot 10^{-3}$ & $312.57$ & $288.33$  \\ 
     & & & &$0.2880$ &  Duan's GARCH & $0.6656$ & $2.2740\cdot 10^{-3}$ & $181.63$ & $780.57$  \\ 
     & & & & & CTS-GARCH & $0.5042$ & $1.3049\cdot 10^{-3}$ & $194.84$ & $586.98$  \\ 
    \hline
    Put & $4219.55$ & $0.0020$ & $4215$ & $0.0360$ &  Duan's GARCH & $-0.2923$ & $2.6660\cdot 10^{-3}$ & $439.87$ & $-45.64$  \\ 
     & & & & & CTS-GARCH & $-0.2196$ & $1.7826\cdot 10^{-3}$ & $389.37$ & $-34.41$  \\ 
     & & & &$0.1480$ &  Duan's GARCH & $-0.4410$ & $2.0529\cdot 10^{-3}$ & $442.75$ & $-288.81$  \\ 
     & & & & & CTS-GARCH & $-0.2956$ & $1.4555\cdot 10^{-3}$ & $283.71$ & $-194.58$  \\ 
     & & & &$0.2880$ &  Duan's GARCH & $-0.4726$ & $1.5977\cdot 10^{-3}$ & $152.80$ & $-613.73$  \\ 
     & & & & & CTS-GARCH & $-0.2807$ & $1.0474\cdot 10^{-3}$ & $178.97$ & $-369.67$  \\
      \hline
    \end{tabular}
    \end{footnotesize}\caption{\label{tbl:DeltaGamma} Greeks for ATM calls and puts on 6/9/2021}
    \end{table}
    
    \begin{table}
    \centering
    \begin{tabular}{ccccccc}
    \hline
     & & \multicolumn{2}{c}{Duan's GARCH Model} & & \multicolumn{2}{c}{CTS-GARCH Model} \\
     \cline{3-4}  \cline{6-7}
    Date & Obs. & MCS (sec) & ANN (sec) & & MCS (sec) & ANN (sec) \\
    \hline
    6/9/2021	& $	558	$ & $	0.1300	$ & $	0.0185	$ & & $	0.3286	$ & $	0.0187	$ \\
    7/7/2021	& $	644	$ & $	0.1387	$ & $	0.0191	$ & & $	0.3411	$ & $	0.0175	$ \\
    8/11/2021	& $	706	$ & $	0.1339	$ & $	0.0169	$ & & $	0.3025	$ & $	0.0172	$ \\
    9/8/2021	& $	729	$ & $	0.1518	$ & $	0.0166	$ & & $	0.3363	$ & $	0.0172	$ \\
    10/6/2021	& $	845	$ & $	0.1422	$ & $	0.0178	$ & & $	0.3390	$ & $	0.0178	$ \\
    11/10/2021	& $	858	$ & $	0.1463	$ & $	0.0183	$ & & $	0.3229	$ & $	0.0175	$ \\
    12/8/2021	& $	843	$ & $	0.1538	$ & $	0.0174	$ & & $	0.3342	$ & $	0.0182	$ \\
    1/12/2022	& $	878	$ & $	0.1598	$ & $	0.0169	$ & & $	0.3356	$ & $	0.0172	$ \\
    2/9/2022	& $	928	$ & $	0.2017	$ & $	0.0193	$ & & $	0.3539	$ & $	0.0207	$ \\
    3/9/2022	& $	923	$ & $	0.1625	$ & $	0.0176	$ & & $	0.3351	$ & $	0.0185	$ \\
    4/6/2022	& $	818	$ & $	0.1410	$ & $	0.0165	$ & & $	0.3285	$ & $	0.0189	$ \\
    5/10/2022	& $	570	$ & $	0.1211	$ & $	0.0163	$ & & $	0.2978	$ & $	0.0171	$ \\
    \hline
    \end{tabular}
    \caption{\label{tbl:timming} Comparison of calculation time using the MCS and ANN methods for Duan's GARCH and CTS-GARCH models}
    \end{table}
    
\section{Conclusion}
\label{Conclusion}
   In this paper, we review Duan's GARCH model and the CTS-GARCH model for option pricing and investigate the use of ANNs to enhance calibration performance. To achieve this goal, we generate training sets for various model parameters and compute calls and puts prices using the MCS method. We then train a three-layer ANN with twenty nodes per layer using the generated training set. Additionally, we create four ANNs for calls and puts under both Duan's GARCH model and the CTS-GARCH model. Finally, we demonstrate the effectiveness of the trained ANNs by using them to calibrate market option prices. The results show that the ANN is significantly faster than the MCS method once it has been trained and it outperforms the MCS method in terms of calibration performance. Furthermore, the ANN method allows us to calculate Greeks, which is not available with the MCS method.

\clearpage
\singlespacing
\bibliographystyle{decsci_mod}
\bibliography{OptionPricingAnnCTSGARCH}
\end{document}